\documentclass[twocolumn,showpacs,amsmath,pra,aps,amssymb,superscriptaddress,nofootinbib]{revtex4-1} 
\usepackage[T1]{fontenc}
\usepackage[latin9]{inputenc}
\usepackage{times}
\usepackage{color} 
\usepackage{xspace}
\usepackage{amssymb,amsmath}
\usepackage{amsbsy}
\usepackage[pdftex]{graphicx}
\usepackage{bm}
\usepackage{float}
\usepackage{adjustbox}
\usepackage{epstopdf}

\usepackage[unicode,breaklinks]{hyperref}
\hypersetup{
    unicode=true,
    plainpages=false, 
    colorlinks=true,
    linkcolor=blue,
    citecolor=blue,
    filecolor=black,
    urlcolor=blue
}
\usepackage{url}
\usepackage{verbatim}

\begin{document}

\title{A damped point-vortex model for polar-core spin vortices in a ferromagnetic spin-1 Bose-Einstein condensate}
\author{Lewis A. Williamson}  
\affiliation{Dodd-Walls Centre for Photonic and Quantum Technologies, Department of Physics, University of Otago, Dunedin 9016, New Zealand}
\affiliation{Physics Department, Lancaster University, Lancaster LA1 4YB, UK} 
\author{P.~B.~Blakie}  
\affiliation{Dodd-Walls Centre for Photonic and Quantum Technologies, Department of Physics, University of Otago, Dunedin 9016, New Zealand}

\begin{abstract}
Ferromagnetic spin-1 Bose-Einstein condensates in the broken-axisymmetric phase support polar-core spin vortices (PCVs), which are intimately linked to the nonequilibrium dynamics of the system. For a purely transversely magnetized system, the Turner point-vortex model predicts that PCVs behave like massive charged particles interacting via a two-dimensional Coulomb potential. We test the accuracy of the Turner model for two oppositely charged PCVs, via comparisons with numerical simulations. While the bare Turner model shows discrepancies with our numerical results, we find that a simple rescaling of the PCV mass gives much better agreement. This can be explained via a phenomenological damping arising from coupling to modes extrinsic to the point-vortex phase space. We also identify the excitations produced following PCV annihilation, which help elucidate recent phase ordering results. We extend the Turner model to cases where the system is magnetized both transversally and axially, identifying a crossover to scalar vortex dynamics for increasing external Zeeman field.
\end{abstract}

\maketitle

\section{Introduction}
Spinor Bose-Einstein condensates (BECs) can exhibit both ferromagnetic and antiferromagnetic order, and possess a rich array of phases with distinct symmetry properties. Associated with these phases are a variety of topological defects and spin textures~\cite{Kawaguchi2012R,StamperKurn2013a}, which play an essential role in nonequilibrium processes such as symmetry breaking and the Kibble-Zurek mechanism~\cite{Kibble1976a,Zurek1985a}, phase ordering dynamics~\cite{Bray1994} and quantum turbulence~\cite{bradley2012,barenghi2001}. Most work on defects and spin textures in spinor condensates have focused on topological and stability aspects~\cite{yip1999,isoshima2001,stoof2001,khawaja2001a,khawaja2001b,zhou2001,mizushima2002,makela2003,ruostekoski2003,semenoff2007,barnett2007,kawaguchi2008,huhtamaki2009,lovegrove2014,borgh2016,tiurev2018,leslie2009b,choi2012,ray2014,hall2016,ollikainen2017,lee2018,ollikainen2019,weiss2019,underwood2020,takeuchi2020}, with less exploration into defect-defect interactions and consequent dynamics. Studies on the interactions and resulting dynamics of half-quantum vortices in antiferromagnetic spin-1 condensates~\cite{eto2011,seo2015,seo2016,kasamatsu2016} and the collisional dynamics of non-abelian vortices in spin-2 condensates~\cite{kobayashi2009,mawson2015}, as well as the dynamics of vortex dipoles across distinct magnetic phases~\cite{kaneda2014}, are notable exceptions, and reveal the rich dynamics possible due to the multicomponent nature of defects in spinor condensates.

The ground state manifold of a ferromagnetic spinor BEC is $\text{SO}(3)$, supporting both nonsingular and singular defects~\cite{lovegrove2014}. Through the use of a quadratic Zeeman field, the spin vector can be constrained to point in a plane transverse to the field direction, and hence has $\text{SO}(2)$ symmetry. This phase is termed the broken-axisymmetric (or easy-plane) phase. In this work we explore the dynamics of polar-core spin vortices (PCVs), which are point defects arising in the broken-axisymmetric phase. These vortices consist of a phase winding of transverse spin density around a polar (i.e.\ unmagnetised) core, and have been observed \emph{in situ} in experiments~\cite{Sadler2006a}. PCVs have been shown to play a crucial role in the symmetry breaking following a quench to the easy-plane phase~\cite{Sadler2006a,Saito2007b}, leading to Kibble-Zurek scaling~\cite{Saito2007a,uhlmann2007,lamacraft2007} and universal long-time phase ordering dynamics~\cite{williamson2016a,williamson2016b,schmied2019,kudo2015}.

In scalar condensates, the point-vortex model~\cite{aref2007} has been shown to describe accurately a plethora of nonequilibrium dynamical effects~\cite{gauthier2019,johnstone2019,sachkou2019,middelkamp2011,navarro2013,moon2015,reeves2013,billam2014,reeves2014,simula2014,billam2015,groszek2018,stockdale2020}. An analogous point-vortex model of PCV dynamics was introduced by Turner~\cite{Turner2009}, where it was argued that the constituent circulations that make up a PCV interact like scalar vortices, but are confined due to spin-exchange interactions. This confining energy manifests itself as a mass, resulting in PCVs behaving like massive charged particles interacting via a two-dimensional Coulomb force. We term this model the ``Turner model''. The presence of a vortex mass makes PCV dynamics vastly different from scalar vortices. With scalar vortices, two like charged vortices circulate around their centroid, while two oppositely charged vortices move in parallel lines (in the absence of damping)~\cite{fetter1965,fetter1966,jones1982}. In contrast, the Turner model predicts that two like charged PCVs should repel, while two oppositely charged PCVs should attract. The qualitative features of the Turner model have been confirmed for two oppositely charged PCVs via simulations of the spin-1 Gross-Pitaevskii equations~\cite{williamson2016c}. However this work also identified quantitative discrepancies between simulations and the Turner model, and hence left questions open regarding the precise quantitative details of the process.

In this work we carry out a detailed comparison between the Turner model of PCV dynamics and numerical simulations of two oppositely charged PCVs. While the bare Turner model shows deviations from numerical simulations, we find that the agreement can be vastly improved by introducing a phenomenological damping of the modes that give rise to the PCV mass, resulting in a simple rescaling of the bare PCV mass. The damping arises due to a coupling of the PCV coordinates to extrinsic modes, resulting in a loss of energy from the point-vortex phase space. We also identify the excitations produced following PCV annihilation, which likely play an important role in the anomalous phase ordering identified in~\cite{williamson2019}. A linear Zeeman field results in the ground state condensate magnetization rotating out of the transverse plane to partially align with the Zeeman field~\cite{murata2007}. We extend the Turner model to PCVs in this phase, finding good agreement with our model and numerical simulations.

The paper is organized as follows. In Sec.~\ref{background} we present background material to understand PCVs and the Turner model. In Sec.~\ref{Turner} we compare the Turner model with numerical simulations for two oppositely charged PCVs, and identify a crucial damping processes missing from the original Turner model. In Sec.~\ref{annihilation} we study the excitations produced following PCV annihilation and in Sec.~\ref{axial} we extend the Turner model to describe systems with both axial and transverse magnetization. In Sec.~\ref{conclusion} we conclude, with various ideas for future work.

\section{Background}\label{background}

\begin{figure}[h]
\includegraphics[trim=0cm 4cm 0cm 4cm,clip=true,width=0.48\textwidth]{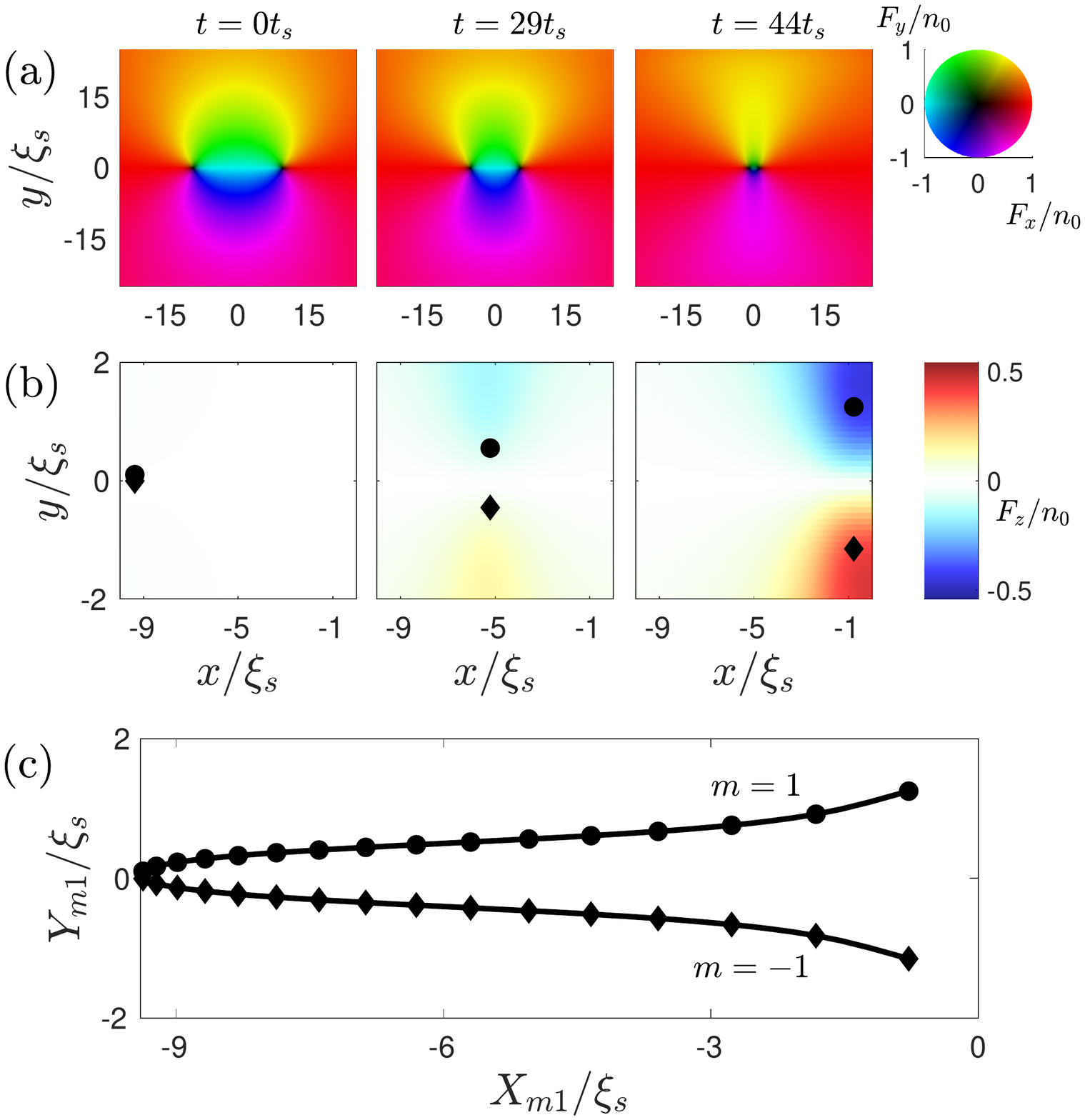}
\caption{\label{fig1}(a) GPE simulations of the dynamics of two oppositely charged PCVs, showing that they attract, as predicted by the Turner model~\eqref{TM}. [The full simulation grid is $100\xi_s\times 100\xi_s$, hence the PCVs are far from the boundary to mitigate boundary effects]. (b) This horizontal motion is caused by a vertical separation of the  $\psi_{\pm 1}$ vortices that constitute a single PCV, such that the stretch energy balances the Magnus force from the horizontal motion. Circles(diamonds) mark the centre of circulation of the vortex in the $\psi_1(\psi_{-1})$ components. The core of the $\psi_1(\psi_{-1})$ vortices are filled by the $\psi_{-1}(\psi_1)$ component, giving rise to an $F_z$ dipole within the core. (c) The coordinates of the $\psi_{\pm 1}$ vortices that constitute the rightward moving (negatively charged) PCV, clearly showing the component separation as the PCV moves. Results are for $q=0.3q_0$.}
\end{figure}

\subsection{Spin-1 BECs}
The atoms in a spin-1 BEC have access to three spin levels $m=-1,0,1$ and can be described by a spinor of three classical fields $\Psi(\mathbf{x})=(\psi_1(\mathbf{x}),\psi_0(\mathbf{x}),\psi_{-1}(\mathbf{x}))^T$. In a flat-bottomed quasi-2D trap, the Gross-Pitaevskii equations (GPEs) are~\cite{Kawaguchi2012R},
\begin{equation}\label{psimdot}
i\hbar\frac{\partial \psi_m}{\partial t}=-\frac{\hbar^2\nabla^2}{2M}\psi_m+\frac{\delta H_\text{int}[\Psi,\Psi^*]}{\delta \psi_m^\dagger},
\end{equation}
with~\cite{Ho1998a,Ohmi1998a,barnett2011}
\begin{equation}\label{Hdef}
\begin{split}
H_\text{int}=\int d^2{\mathbf{x}}\,\bigg[&\frac{g_n}{2}n(\mathbf{x})^2+\frac{g_s}{2}\left|\mathbf{F}(\mathbf{x})\right|^2\\&-p\Psi^\dagger(\mathbf{x})f_z\Psi(\mathbf{x})+q\Psi^\dagger(\mathbf{x})f_z^2\Psi(\mathbf{x})
\bigg].
\end{split}
\end{equation}
The spin-1 atoms interact via spin-independent (strength $g_n$) and spin-dependent (strength $g_s$) interactions, with $n(\mathbf{x})=\Psi^\dagger(\mathbf{x})\Psi(\mathbf{x})$ the areal density and $\mathbf{F}(\mathbf{x})=\sum_{\mu=x,y,z}\Psi^\dagger(\mathbf{x}) f_\mu\Psi(\mathbf{x})\hat{\mathbf{s}}_\mu$ the areal spin density for spin-1 Pauli matrices $f_\mu$ and spin directions $\hat{\mathbf{s}}_\mu$. We consider ferromagnetic ($g_s<0$) spin interactions, arising for example in $^{87}$Rb~\cite{schmaljohann2004,Chang2004a} and $^7$Li~\cite{huh2020}, which favours a nonzero magnetization density. A magnetic field along $\hat{\mathbf{s}}_z$, along with microwave dressing techniques~\cite{gerbier2006,leslie2009}, results in a Zeeman splitting of the spin levels, and has both a linear $p^\prime$ and quadratic $q$ contribution. The linear Zeeman shift can be combined with the Lagrange multiplier $\lambda$ that enforces conservation of total axial magnetization and hence $p=p^\prime+\lambda$~\cite{Stenger1998a}. For $0<q<q_0= 2|g_s|n_0$ ($n_0$ is the mean condensate density) and $|p|<q$ the ground state is in the broken-axisymmetry (BA) phase, whereby the quadratic Zeeman energy favours a transverse magnetization that breaks the axial symmetry of the Hamiltonian~\cite{Stenger1998a,murata2007}. Here $q_0$ is a quantum critical point separating the $p=0$ BA phase from from the polar [$\Psi=(0,\psi_0,0)^T]$ phase.

\subsection{Polar-core spin vortices}
For $p=0$ in the BA phase, the quadratic Zeeman field confines the spin to point entirely in the $(\hat{\mathbf{s}}_x,\hat{\mathbf{s}}_y)$ plane and the ground state has only transverse magnetization. The ground state can be parameterized as~\cite{murata2007}
\begin{equation}\label{EPgs}
\Psi=e^{i\theta}e^{if_z\varphi}\sqrt{\frac{n_0}{2}}\left(\begin{array}{c}\sin\beta\\\sqrt{2}\cos\beta \\\sin\beta\end{array}\right),
\end{equation}
with $\text{U}(1)$ symmetries due to global phase rotations $e^{i\theta}$ and spin rotations $e^{-if_z\theta}$ around $\hat{\mathbf{s}}_z$. Here $\cos (2\beta)=q/q_0$. The resulting spin density is
\begin{equation}
\mathbf{F}=n_0\sqrt{1-\left(\frac{q}{q_0}\right)^2}(\cos\varphi,\sin\varphi,0).
\end{equation}
A $2\pi\kappa$ ($\kappa\in\mathbb{Z}\setminus\{0\}$) phase winding of the transverse spin $F_\perp=F_x+iF_y$ results when $\varphi=\kappa\phi(\mathbf{x})$, with $\phi(\mathbf{x})=\operatorname{phase}(x+iy)$ the phase of the complex number $x+iy$. When there is no additional global phase rotation \footnote{A combination of spin and global phase rotations in Eq.~\eqref{EPgs} produces a Mermin-Ho vortex~\cite{mizushima2002}, which also gives rise to a phase winding of the transverse spin, but will otherwise have different properties to the PCVs explored here.}, the state is a PCV~\cite{Kawaguchi2012R},
\begin{equation}\label{ansatz}
\Psi({\mathbf{x}})\approx \sqrt{\frac{n_0}{2}}\left(\begin{array}{c}\sin\beta e^{-i\kappa\phi(\mathbf{x})}\\\sqrt{2}\cos\beta \\\sin\beta e^{i\kappa\phi(\mathbf{x})}\end{array}\right).
\end{equation}
This PCV state is approximate as in reality a core will form, to avoid the divergences of $\nabla\psi_{\pm 1}$. Hence the density and potentially the phase profiles will be modified for distances $|\mathbf{x}|\lesssim \xi_s$, with $\xi_s= \hbar/\sqrt{Mq_0}$ the spin healing length. Within the core there is a peak in the occupation of the $\psi_0$ (polar) component, hence the terminology ``polar-core spin vortex''. As we will see, the core structure plays an integral role in the PCV motion, drastically changing the dynamics compared to the scalar case. Since the $\psi_{\pm 1}$ components of a single PCV circulate in opposite directions, the velocity field generated by a collection of PCVs consists of counter-flowing $\psi_1$ and $\psi_{-1}$ currents. Hence the flow field around a PCV transports no mass but will transport axial magnetization $F_z=|\psi_1|^2-|\psi_{-1}|^2$~\cite{Yukawa2012}. 

\subsection{Scalar point-vortex model}
In a scalar BEC, the point-vortex model is derived by assuming that the superfluid can be described by a collection of vortex coordinates $\mathbf{X}_k$ that give the points of $\delta$-function divergence of the vorticity of the superfluid velocity field. One then stipulates a vortex ansatz for the condensate field, $\psi(\mathbf{x},t)=\sqrt{n_0}\prod_ke^{i\kappa_k\phi[\mathbf{x}-\mathbf{X}_k(t)]}$ ($\kappa_k\in\mathbb{Z}\setminus\{0\}$ are the vortex charges), which evolves according to the scalar Gross-Pitaevskii equation. Deriving equations of motion for the vortex positions $\mathbf{X}_k(t)$ using either variational Lagrangian~\cite{lucas2014} or hydrodynamic~\cite{lin1999} techniques gives the point-vortex equations of motion,
\begin{equation}\label{scalarvortex}
\hat{\mathbf{z}}\times\dot{\mathbf{X}}_k=-\sum_{j\ne k}\frac{\hbar \kappa_j}{M}\frac{\mathbf{X}_k-\mathbf{X}_j}{|\mathbf{X}_k-\mathbf{X}_j|^2}.
\end{equation}
Equation~\eqref{scalarvortex} shows that a scalar vortex follows the velocity field generated by the remaining vortices. This is necessary to conserve the fluid momentum in the presence of no external forces, in analogy to the Magnus effect in classical fluid dynamics~\cite{thouless1996,ao1993,sonin1997,simula2018}.

Two fundamental assumptions of the scalar point-vortex model are that interactions with sound waves can be ignored and that the precise core features of the vortices do not affect the mesoscopic dynamics of the system, hence the use of $\delta$-function cores suffice. In contrast, we will show that the core features of a PCV have a radical effect on the mesoscopic dynamics of the system and that coupling to spin waves is non-negligible. Despite this, a point-vortex model is still extremely useful at describing the bulk dynamics of PCVs.

\subsection{PCV point-vortex model}
The variational Lagrangian method to derive the point-vortex model for scalar vortices, Eq.~\eqref{scalarvortex}, can be adapted to PCVs~\cite{Turner2009,williamson2016c}. We start with an ansatz consisting of a product of PCVs~\cite{williamson2016c},
\begin{equation}\label{ansatz2}
\Psi({\mathbf{x}},t)=\sqrt{\frac{n_0}{2}}\left(\begin{array}{c}\sin\beta\prod_k g_{1k}(\mathbf{x},t)e^{-i\kappa_k \phi(\mathbf{x}-\mathbf{X}_{1k})}\\\prod_k\sqrt{2}\cos\beta g_{0k}(\mathbf{x},t)\\\sin\beta\prod_k g_{-1k}(\mathbf{x},t)e^{i\kappa_k \phi(\mathbf{x}-\mathbf{X}_{-1k})}\end{array}\right),
\end{equation}
with $\mathbf{X}_{mk}(t)$ the centre of a vortex in spin component $m$, $\kappa_k$ the PCV charge and $g_{mk}(\mathbf{x},t)$ the amplitude profiles giving rise to the PCV core. The $\mathbf{X}_{1k}$ and $\mathbf{X}_{-1k}$ coordinates that constitute a single PCV will tend to move in opposite directions, due to the counter-flowing $\psi_{\pm 1}$ fields generated by other PCVs and Eq.~\eqref{scalarvortex}. This ``stretching'' is restricted, ultimately by the spin exchange term in $g_s|\mathbf{F}|^2$, which favours the phase profiles of $\psi_1$ and $\psi_{-1}$ to overlap. In addition, other energy terms arising from overlapping vortex cores in the $\psi_{\pm 1}$ components may restrict the stretching. Hence the additional energy term $H_\text{int}[\Psi,\Psi^\dagger]$ must be considered in the PCV dynamics.

The ansatz~\eqref{ansatz2}, as it currently stands, allows for coupling between phase and amplitude excitations via coupling between $\mathbf{X}_{\pm 1k}$ and excitations of the amplitude profiles $g_{mk}$. To obtain a point-vortex description of PCV dynamics, we make the approximation that the core profiles $g_{mk}$ are stationary, depending only on $t$ through the vortex coordinates $\mathbf{X}_{mk}$, $g_{mk}(\mathbf{x},t)\rightarrow g_m(\mathbf{x},\mathbf{X}_{1k},\mathbf{X}_{-1k})$. As a corollary, $H_\text{int}[\Psi,\Psi^\dagger]$ depends only on the vortex coordinates $\mathbf{X}_{mk}$. Assuming also that the contribution of each PCV to $H_\text{int}$ depends only on the magnitude of the PCV stretch $|\mathbf{X}_{1k}-\mathbf{X}_{-1k}|$ allows one to further decompose $H_\text{int}$ as
\begin{equation}\label{Hint}
H_\text{int}[\Psi,\Psi^\dagger]\approx \frac{\pi\hbar^2 n_0\sin^2\beta}{4M}\sum_k \kappa_k^2 u(|\mathbf{X}_{1k}-\mathbf{X}_{-1k}|)
\end{equation}
for individual PCV stretch energies $\propto u(|\mathbf{X}_{1k}-\mathbf{X}_{-1k}|)$ (the prefactor has been chosen to simplify the equations of motion below). Adding this to the scalar vortex interaction from Eq.~\eqref{scalarvortex}, assuming that the PCV stretching remains small compared to the distance between PCVs, and neglecting the kinetic energy arising from the vortex cores, gives~\cite{Turner2009,williamson2016c}
\begin{equation}\label{TMi}
\hat{\mathbf{z}}\times\dot{\mathbf{X}}_{mk}=\sum_{j\ne k}\frac{\hbar m\kappa_j}{M}\frac{\mathbf{X}_{mk}-\mathbf{X}_{mj}}{|\mathbf{X}_{mk}-\mathbf{X}_{mj}|^2}-\frac{m\hbar\kappa_k}{2M}\nabla_{\mathbf{X}_{mk}}u,
\end{equation}
with $\nabla_\mathbf{u}=\hat{\mathbf{x}}\partial/\partial u_x+\hat{\mathbf{y}}\partial/\partial u_y$ the gradient derivative with respect to a vector $\mathbf{u}=(u_x,u_y)$. We see that the fluid momentum in component $m$ is now balanced by the force $\propto -\nabla_{\mathbf{X}_{mk}}u$~\cite{Turner2009}. We introduce position $\mathbf{R}_k$ and stretch $\mathbf{r}_k$ coordinates for each PCV,
\begin{equation}
\begin{split}
\mathbf{R}_k=(X_k,Y_k)=&\frac{{\mathbf{X}}_{1k}+{\mathbf{X}}_{-1k}}{2},\\
\mathbf{r}_k=(x_k,y_k)=&{\mathbf{X}}_{1k}-{\mathbf{X}}_{-1k}.
\end{split}
\end{equation}
and hence rewrite Eqs.~\eqref{TMi} as
\begin{subequations}\label{TM}
\begin{align}
\hat{\mathbf{z}}\times\dot{\mathbf{R}}_k=&-\nabla_{\mathbf{r}_k}\frac{\hbar\kappa_k}{2M}u(|\mathbf{r}_k|)\label{TM1}\\
\hat{\mathbf{z}}\times\dot{\mathbf{r}}_k=&\sum_{j\ne k}\frac{2\hbar \kappa_j}{M}\frac{\mathbf{R}_k-\mathbf{R}_j}{|\mathbf{R}_k-\mathbf{R}_j|^2}.\label{TM2}
\end{align}
\end{subequations}
Equation~\eqref{TM2} can be written in terms of the gradient of a two-dimensional Coulomb potential $\propto -\ln |(\mathbf{R}_k-\mathbf{R}_j)/\xi_s |$. Hence Eqs.~\eqref{TM} both take the form of a Magnus force balanced by the gradient of a potential.

The transverse spin will be zero at $\mathbf{x}=\mathbf{R}_k$ to avoid a singularity in the phase winding of $\varphi$. Equation~\eqref{TM2} predicts that the $\psi_{\pm 1}$ flow fields will cause the PCV to stretch. This results in a force on the $\psi_{\pm 1}$ components, arising from the stretch energy~\eqref{Hint}, with the net effect that two PCVs of opposite(same) charge attract(repel), Eq.~\eqref{TM1}. The equations of motion~\eqref{TM} are derived from a time translationally invariant and hence energy conserving Hamiltonian, with a conserved dimensionless PCV energy~\cite{williamson2016c},
\begin{equation}\label{PCVenergy}
\mathcal{H}=\mathcal{H}_R+\mathcal{H}_r
\end{equation}
with
\begin{subequations}
\begin{align}
\mathcal{H}_R=&-2\sum_{j,k>j}\kappa_k\kappa_j\ln \left|\frac{\mathbf{R}_k-\mathbf{R}_j}{\xi_s}\right|^2\\
\mathcal{H}_r=&\sum_k \kappa_k^2 u(|\mathbf{r}_k|)\label{Hintstr}.
\end{align}
\end{subequations}

(The condensate energy is related to Eq.~\eqref{PCVenergy} by scaling by the prefactor in Eq.~\eqref{Hint}, which includes the factor of $\pi$ often included in the scalar point-vortex energy.)

The assumption that the profiles $g_{mk}$ depend only on time through the coordinates $\mathbf{X}_{mk}$ removes the possibility of coupling to dynamical degrees of freedom outside of the $\{\mathbf{X}_{1k},\mathbf{X}_{-1k}\}$ phase space (or equivalently the $\{\mathbf{R}_k,\mathbf{r}_k\}$ phase space). For scalar vortices, this decoupling is a reasonable approximation in many cases~\cite{lucas2014}. As we will later show, it is in fact unreasonable for PCVs. However, a simple linear damping term in Eq.~\eqref{TM2} allows for the effects of additional modes to be included, leading to an accurate point-vortex model for the coordinates $\mathbf{R}_k$.

\section{Testing the Turner model of PCV dynamics}\label{Turner}

We analyze the dynamics of two oppositely charged PCVs with centres of circulation at positions $\mathbf{R}_1(0)=(-10\xi_s,0)$ (charge $\kappa_1=-1$) and $\mathbf{R}_2(0)=(10\xi_s,0)$ (charge $\kappa_2=1$), see Fig.~\ref{fig1}(a). The initial two-PCV state is,
\begin{equation}\label{PCVi}
\Psi({\mathbf{x}})=\sqrt{\frac{n_0}{2}}\left(\begin{array}{c}\sin\beta e^{i\phi(\mathbf{x}-\mathbf{R}_1(0))}e^{-i\phi(\mathbf{x}-\mathbf{R}_2(0))}\\\sqrt{2}\cos\beta\\\sin\beta e^{-i\phi(\mathbf{x}-\mathbf{R}_1(0))}e^{i\phi(\mathbf{x}-\mathbf{R}_2(0))}\end{array}\right),
\end{equation}
which we evolve using the spin-1 GPEs~\eqref{psimdot}. For times $-10t_s\le t<0$ ($t_s= \hbar/q_0$) we include in Eq.~\eqref{psimdot} an energy damping term of strength $\gamma$ \footnote{This is done by solving $i\hbar\dot{\psi}_m=(1-i\gamma)(\mathcal{L}_m[\psi_m]-(\mu_m+mp)\psi_m)$, where $\gamma$ is the dimensionless damping rate, $\mathcal{L}_m$ is the conservative time evolution from Eq.~\eqref{psimdot} and $\mu_m$ is the uniform ground-state chemical potential~\cite{Kawaguchi2012R}. We set $\gamma=1$.}. This allows a core structure to develop, while having only a small affect on $|\mathbf{R}_1-\mathbf{R}_2|$ as long as $\gamma\lesssim 1$. For times $t\ge 0$ we set $\gamma=0$. An example case of PCV dynamics obtained from our simulations is shown in Fig.~\ref{fig1}. As predicted by Eqs.~\eqref{TM}, the oppositely charged PCVs attract. Within each PCV, the $\psi_{\pm 1}$ vortex cores separate, along a line orthogonal to $\mathbf{R}_2-\mathbf{R}_1$, see Fig.~\ref{fig1}(b),(c).

[In all simulations we use $n_0=10^4\xi_s$ and $g_n=10|g_s|$. We solve Eq.~\eqref{psimdot} on a $N=1024\times 1024$ grid with physical size $100\xi_s\times 100\xi_s$ using a fourth order Runge-Kutta method (time step $=0.002 t_s$) with periodic boundary conditions and kinetic energy operator evaluated to spectral accuracy. The PCVs remain far from the boundary, which mitigates the effect of image charges and an initial phase discontinuity along $\mathbf{x}=(x,\pm 100\xi_s)$. To detect vortices, we interpolate the wavefunction onto a $100\times$denser grid in a region around each vortex and find points of diverging vorticity $\nabla\times\nabla \operatorname{phase}(\psi_{\pm 1})$.]

\subsection{PCV position dynamics}

\begin{figure}
\includegraphics[trim=0cm 3cm 0cm 5cm,clip=true,width=0.48\textwidth]{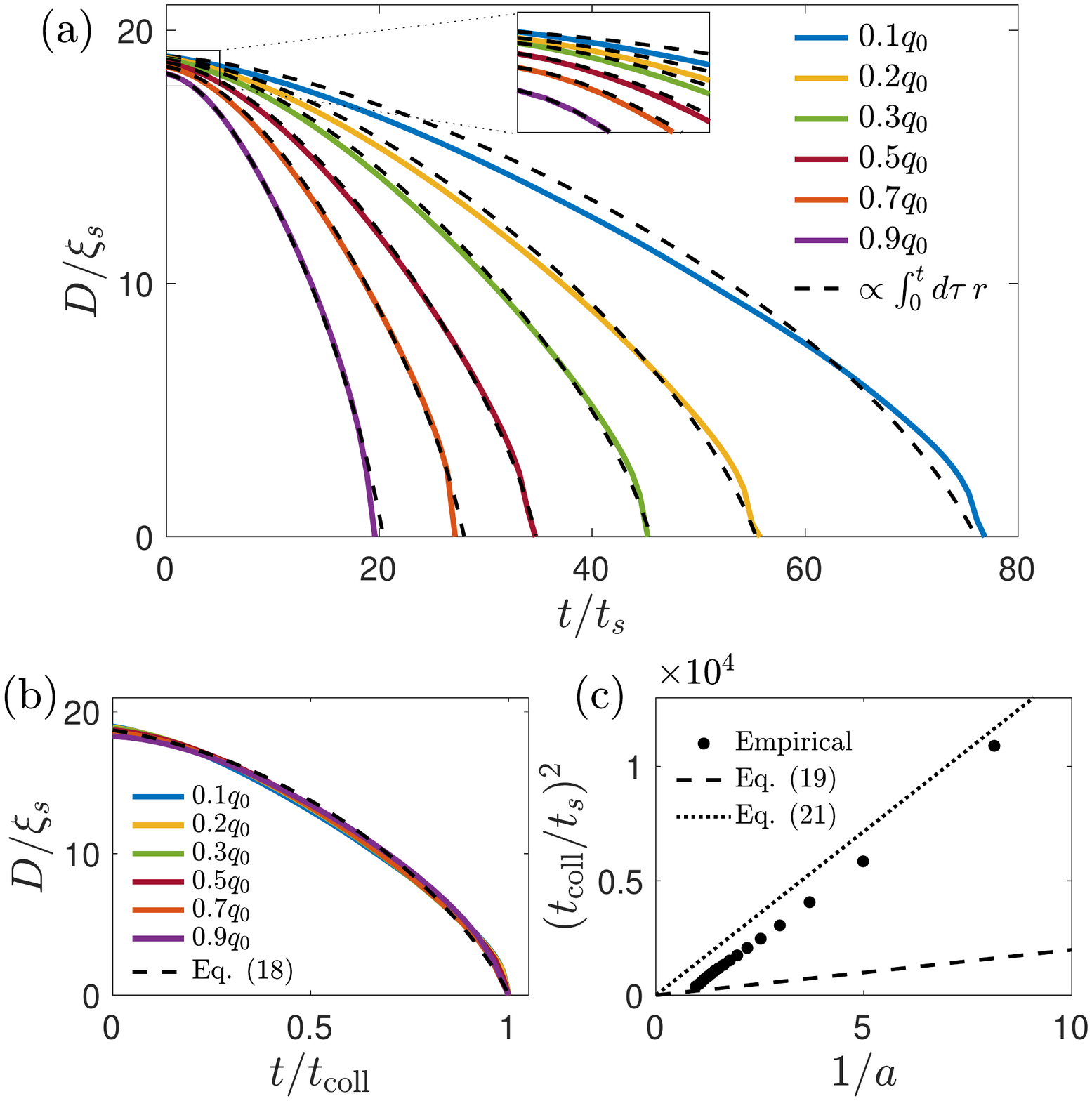}
\caption{\label{fig2}(a) The PCV separation $D(t)$ (colored solid curves) decreases proportional to $\int_0^t r(\tau)\,d\tau$ (matching black dashed curves), as predicted by the Turner model with a harmonic stretch energy, for $q/q_0=(0.1,0.2,0.3,0.5,0.7,0.9)$ (blue, yellow, green, red, orange, purple). The enlarged region shows the initial PCV velocity. (b) The solid curves from (a) collapse onto a single functional form when time is rescaled by the PCV collision time $t_\text{coll}$. This curve is well described by the analytic prediction~\eqref{Dann} (black dashed curve) with $t_\text{coll}$ in replace of $t_\text{coll}^\text{ann}$. (d) The PCV collision time (black dots) scales as $(t_\text{coll})^2\propto 1/a$, as predicted by Eq.~\eqref{tann}. However, the analytic prediction Eq.~\eqref{tann} (black dashed line) underestimates $t_\text{coll}$ and hence a rescaling factor is needed, Eq.~\eqref{tcoll2}, (black dotted line).}
\end{figure}

Solving the Turner model~\eqref{TM} for the PCV coordinates $\mathbf{R}_k$ and $\mathbf{r}_k$ requires stipulating a form for the potential $u(r_k)$. We find that a harmonic potential
\begin{equation}\label{uapprox}
u(r_k)=\frac{1}{2}\frac{a}{\xi_s^2}r_k^2
\end{equation} 
describes the dynamics very well, with $a$ a fitted ``spring constant''. Taking a second time derivative of Eq.~\eqref{TM1} and using Eqs.~\eqref{TM2} and~\eqref{uapprox} gives
\begin{equation}\label{coulomb}
m_v\ddot{\mathbf{R}}_k=\frac{\pi\hbar^2 n_0\sin^2\beta}{M}\sum_{j\ne k}\kappa_j\kappa_k\frac{\mathbf{R}_k-\mathbf{R}_j}{|\mathbf{R}_k-\mathbf{R}_j|^2}
\end{equation}
with,
\begin{equation}
m_v=\frac{\pi M n_0\xi_s^2\sin^2\beta}{a}.
\end{equation}
The prefactors in Eq.~\eqref{coulomb} are such that the right-hand side is the gradient (with respect to $\mathbf{R}_k$) of the condensate kinetic energy. Equation~\eqref{coulomb} predicts that PCVs behave like massive charged particles moving under the influence of the two-dimensional Coulomb interaction \footnote{PCVs of charge $|\kappa_k|>1$ decay into PCVs of lower charge, which repel due to the repulsion of like-charged PCVs predicted by Eq.~\eqref{coulomb}.}.

In Fig.~\ref{fig2}(a) we show simulation results for the PCV separation $D(t)=|\mathbf{R}_2(t)-\mathbf{R}_1(t)|$ for different $q$ values. We test Eq.~\eqref{TM1}, assuming a potential~\eqref{uapprox}, by comparing $D(t)$ with the integral $2a\int_0^t d\tau\,r(\tau)/t_s$. Here $r=|\mathbf{r}_1|=|\mathbf{r}_2|$ is the stretching of either PCV, which we obtain from the GPE simulations. Equation~\eqref{TM1} is satisfied accurately for $q\gtrsim 0.1q_0$, with $a$ as a fitting parameter. For small $q$, the full $\text{SO}(3)$ spin manifold will become accessible and hence we expect deviation from the Turner model \footnote{The fit for $0<q\lesssim 0.1q_0$ can be improved by adding a term $\propto r_k$ in the potential~\eqref{uapprox}.}. 

Equation~\eqref{coulomb} can be solved analytically for the two-PCV setup in Fig.~\ref{fig1} to give,
\begin{equation}\label{Dann}
D(t)=D(0)\frac{\exp\left\{-\left[\operatorname{erf}^{-1}\left(\frac{(1-b)t}{t_\text{coll}^\text{ann}}+b\right)\right]^2\right\}}{\exp\left\{-\left[\operatorname{erf}^{-1}b\right]^2\right\}}.
\end{equation}
Here
\begin{equation}\label{tann}
\begin{split}
t_\text{coll}^\text{ann}=&\sqrt{\frac{\pi}{a}}\frac{D(0)}{2c_s}(1-b)\exp\left\{\left[\operatorname{erf}^{-1}b\right]^2\right\}\\
\propto &\sqrt{m_v}+\mathcal{O}(b)
\end{split}
\end{equation}
is the analytic prediction for the PCV collision time $t_\text{coll}$, with $c_s=\xi_s/t_s$ the characteristic spin-wave speed, and
\begin{equation}\label{bdef}
b=\operatorname{erf}\left(-\frac{\dot{D}(0)}{2c_s\sqrt{a}}\right)
\end{equation}
allows for a nonzero initial PCV velocity $\dot{D}(0)$, which is present in our simulation due to the initial damped evolution, see enlarged region in Fig.~\ref{fig2}(a). We can easily extract the PCV collision time $t_\text{coll}$ from our GPE simulations. Rescaling time by $t_\text{coll}$ results in the curves for different $q$ values from Fig.~\ref{fig2}(a) reducing to the same functional form, see Fig.~\ref{fig2}(b). Hence the entire $q$ dependence of the dynamics of $D$ is contained in the parameter $t_\text{coll}$, as predicted by Eq.~\eqref{Dann}. The agreement with Eq.~\eqref{Dann} is excellent with $t_\text{coll}^\text{ann}$ replaced by $t_\text{coll}$ and $b\approx 0.17$. We find that the predicted scaling of $(t_\text{coll})^2\propto 1/a$ from Eq.~\eqref{tann} (to zeroth order in $\dot{D}(0)$) is also very well satisfied, see Fig.~\ref{fig2}(c). In addition, we find that $t_\text{coll}\propto D(0)$ (not shown), as predicted by Eq.~\eqref{tann}. The initial PCV velocity, obtained from Eq.~\eqref{bdef} using $b=0.17$, is $\dot{D}(0)\approx 0.3\sqrt{a}$, and decreases as $q$ is increased, see enlarged region in Fig.~\ref{fig2}(a).

Despite these consistencies, Eq.~\eqref{tann} underestimates the collision time, see Fig.~\ref{fig2}(c). To get improved agreement, we need to rescale Eq.~\eqref{tann} by a scaling factor $A$,
\begin{equation}\label{tcoll2}
t_\text{coll}^\text{ann}\rightarrow \tau_\text{coll}^\text{ann}=At_\text{coll}^\text{ann}\approx 2.7t_\text{coll}^\text{ann}.
\end{equation}
This can be interpreted as an enhancement of the PCV mass to the effective mass
\begin{equation}\label{effmass}
m_v^\text{eff}=A^2 m_v\approx 7m_v.
\end{equation}
The necessity of using an effective mass, rather than the bare PCV mass, is the first evidence of a coupling to extrinsic degrees of freedom, outside of the $\left\{\mathbf{R}_k,\mathbf{r}_k\right\}$ phase space. We will provide more evidence of this shortly, where we will see that the scaling factor $A$ can be written in terms of a damping rate of the coordinates $\mathbf{r}_k$. The notion of an effective mass arising from a coupling to extrinsic modes is reminiscent of the notion of electron effective mass in solids~\cite{kittel2004}. Note that we see no $q$ dependence on the enhancement factor in Eq.~\eqref{effmass}, hence the $q$ dependence is entirely contained within the bare mass $m_v$.

\subsection{PCV stretch dynamics}

\begin{figure}
\includegraphics[trim=0cm 5cm 0cm 5.5cm,clip=true,width=0.48\textwidth]{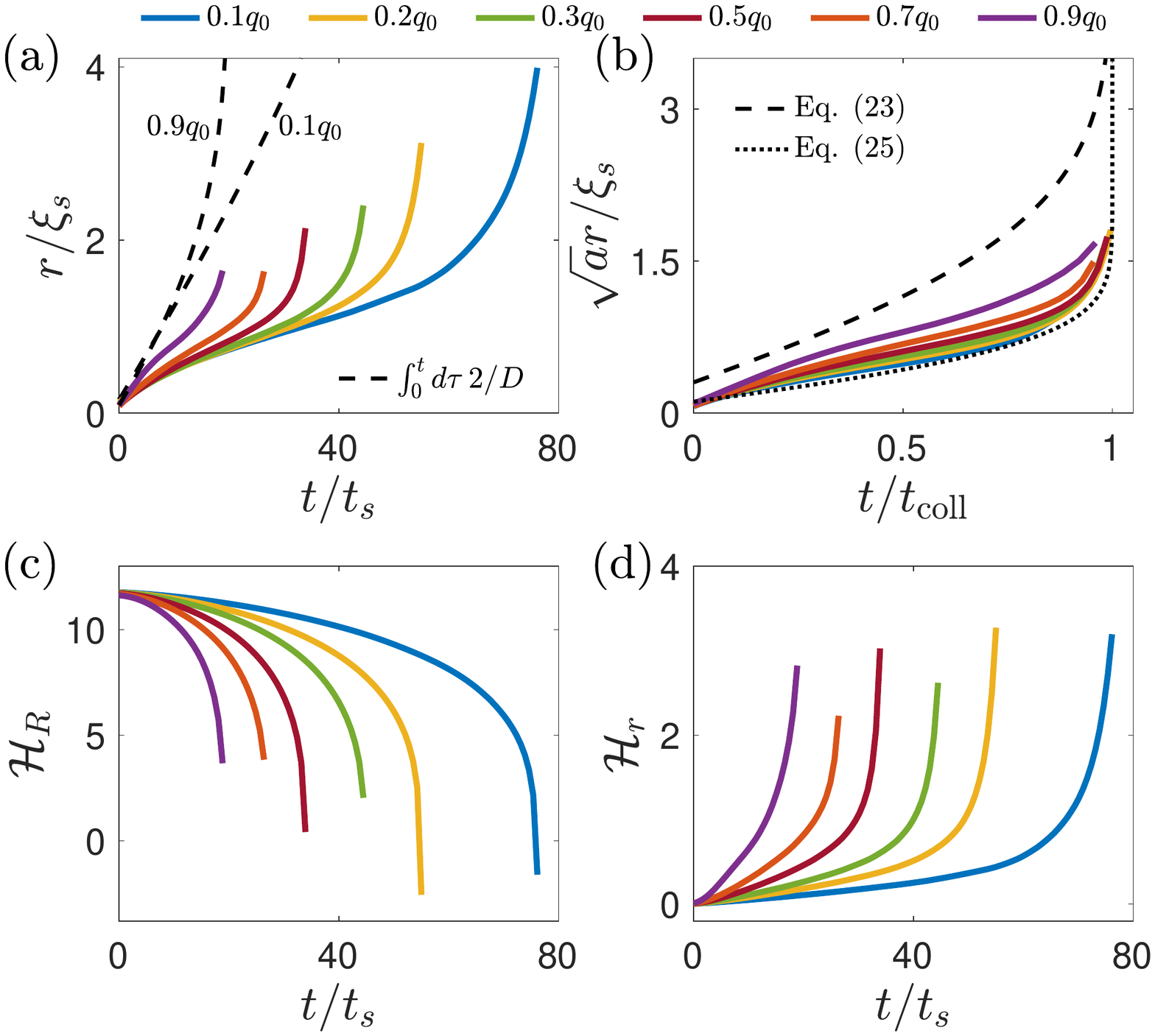}
\caption{\label{fig3}(a) The stretching of the two PCVs increases with time (colored solid curves) for $q/q_0=(0.1,0.2,0.3,0.5,0.7,0.9)$ (blue, yellow, green, red, orange, purple). The corresponding Turner model predictions, obtained from integrating Eq.~\eqref{TM2}, are much larger, ranging from the lower black dashed curve curve ($q=0.1q_0$) to the upper black dashed curve ($q=0.9q_0$). (b) Rescaling $r$ by $\sqrt{a}$ and $t$ by $t_\text{coll}$ does not result in the curves from (a) reducing to a single curve. The analytic result Eq.~\eqref{rann} (black dashed line) gives a much larger stretching than that observed for any $q$ value. The agreement with the small $q$ curves is improved with a phenomenologically damping of $\dot{r}$ to account for energy lost to other modes (black dotted line). (c),(d) As the PCVs come together, the energy $\mathcal{H}_R$ decreases much more than the stretch energy $\mathcal{H}_r$ increases, hence there is an overall loss of $\mathcal{H}$ from the system. [Note the vertical axis scale in (d) is 4 $\times$ smaller than in (c).]}
\end{figure}

We now turn to the PCV stretch coordinate $r$. In Fig.~\ref{fig3}(a) we show simulation results for $r(t)$ for different $q$ values. We test Eq.~\eqref{TM2} by comparing $r(t)$ with the integral $\int_0^t d\tau\,2/D(\tau)$ for the two extreme cases $q=0.1q_0$ and $q=0.9q_0$ (the remaining $\int_0^t d\tau\,2/D(\tau)$ results fall between these two curves). Evidently, the stretch coordinate $r$ is much smaller than the prediction of Eq.~\eqref{TM2}. Note that the oscillation of $r$ observed in~\cite{williamson2016c} is not present in Fig.~\ref{fig3}(a). This is due to the initial damped evolution, which removes the energy liberated when the core forms and the resulting long-lived oscillation.

The analytic prediction for $r$, obtained from solving Eqs.~\eqref{TM} with the potential~\eqref{uapprox}, is
\begin{equation}\label{rann}
r(t)=\frac{2\xi_s}{\sqrt{a}}\operatorname{erf}^{-1}\left(\frac{(1-b)t}{t_\text{coll}^\text{ann}}+b\right).
\end{equation}
Results for $\sqrt{a}r(t)$ versus $t/t_\text{coll}$ are shown in Fig.~\ref{fig3}(b), where we see that the different $\sqrt{a}r(t)$ show poor collapse onto a single functional form (in contrary to the prediction of Eq.~\eqref{rann}). Furthermore the analytic prediction~\eqref{rann} (with $t_\text{coll}^\text{ann}\rightarrow t_\text{coll}$) overestimates the stretching. This suggests that energy from the Coulomb potential $\mathcal{H}_R$ is not being entirely transferred into the energy term $\mathcal{H}_r$, in contrary to the conservation law Eq.~\eqref{PCVenergy}.

In Fig.~\ref{fig3}(c),(d) we plot the two energy terms from Eq.~\eqref{PCVenergy} using the simulation results for $\mathbf{R}_k$ and $\mathbf{r}_k$, with the potential~\eqref{uapprox} for $\mathcal{H}_r$ and $a$ obtained from the fits in Fig.~\ref{fig2}(a). The PCV energy $\mathcal{H}_R$ decreases as the PCVs collide, Fig.~\ref{fig3}(c) while the energy $\mathcal{H}_r$ increases, Fig.~\ref{fig3}(d). However, the loss in $\mathcal{H}_R$ is approximately four times greater than the gain in $\mathcal{H}_r$ and hence most of the liberated $\mathcal{H}_R$ energy is lost to other excitations in the system, rather than coupling to the PCV dynamical variable $r$.

\subsection{The damped Turner model}\label{dampedTurner}
We now present a simple modification to Eq.~\eqref{TM2} that gives rise to the effective mass, Eq.~\eqref{effmass}, and which also improves the agreement in Fig.~\ref{fig3}(b). Motivated by the loss of energy identified in Fig.~\ref{fig3}(c),(d), we introduce a phenomenologically damping term $-\Gamma\hat{\mathbf{z}}\times\dot{\mathbf{r}}_k$ on the right-hand side of Eq.~\eqref{TM2}. Solving Eq.~\eqref{TM1} and the modified Eq.~\eqref{TM2} gives the same solution~\eqref{Dann} but with the modified PCV collision time Eq.~\eqref{tcoll2}. The scaling factor $A$ is related to the damping rate $\Gamma$ via,
\begin{equation}
A=\sqrt{1+\Gamma}.
\end{equation}
Meanwhile Eq.~\eqref{rann} is modified to
\begin{equation}\label{rann3}
r(t)=\frac{2\xi_s}{1+\Gamma}\sqrt{\frac{1+\Gamma}{a}}\operatorname{erf}^{-1}\left(\frac{(1-b)t}{\tau_\text{coll}}+b\right).
\end{equation}
Using the effective mass Eq.~\eqref{effmass} gives $\Gamma\approx 6$, which when used in Eq.~\eqref{rann3} gives good agreement with the $r$ curve of the smallest $q$ value, see Fig.~\ref{fig3}(b). Note Fig.~\ref{fig3}(b) suggests a $q$ dependence on the scaling~\eqref{rann3} that is not accounted for by our simple damping model. A more accurate determination of this damping process is an interesting topic for future work. The effective mass, however, appears insensitive to this $q$ dependence and hence our phenomenological damping theory describes very well the behaviour in Fig.~\ref{fig2}. Note also that despite the addition of this damping and the new parameter $\Gamma$, the effect on the dynamics of $D$ is to simply modify the PCV collision time.

\subsection{Probing the PCV stretch energy}\label{energy}
\begin{figure}
\includegraphics[trim=0cm 4cm 0cm 7cm,clip=true,width=0.48\textwidth]{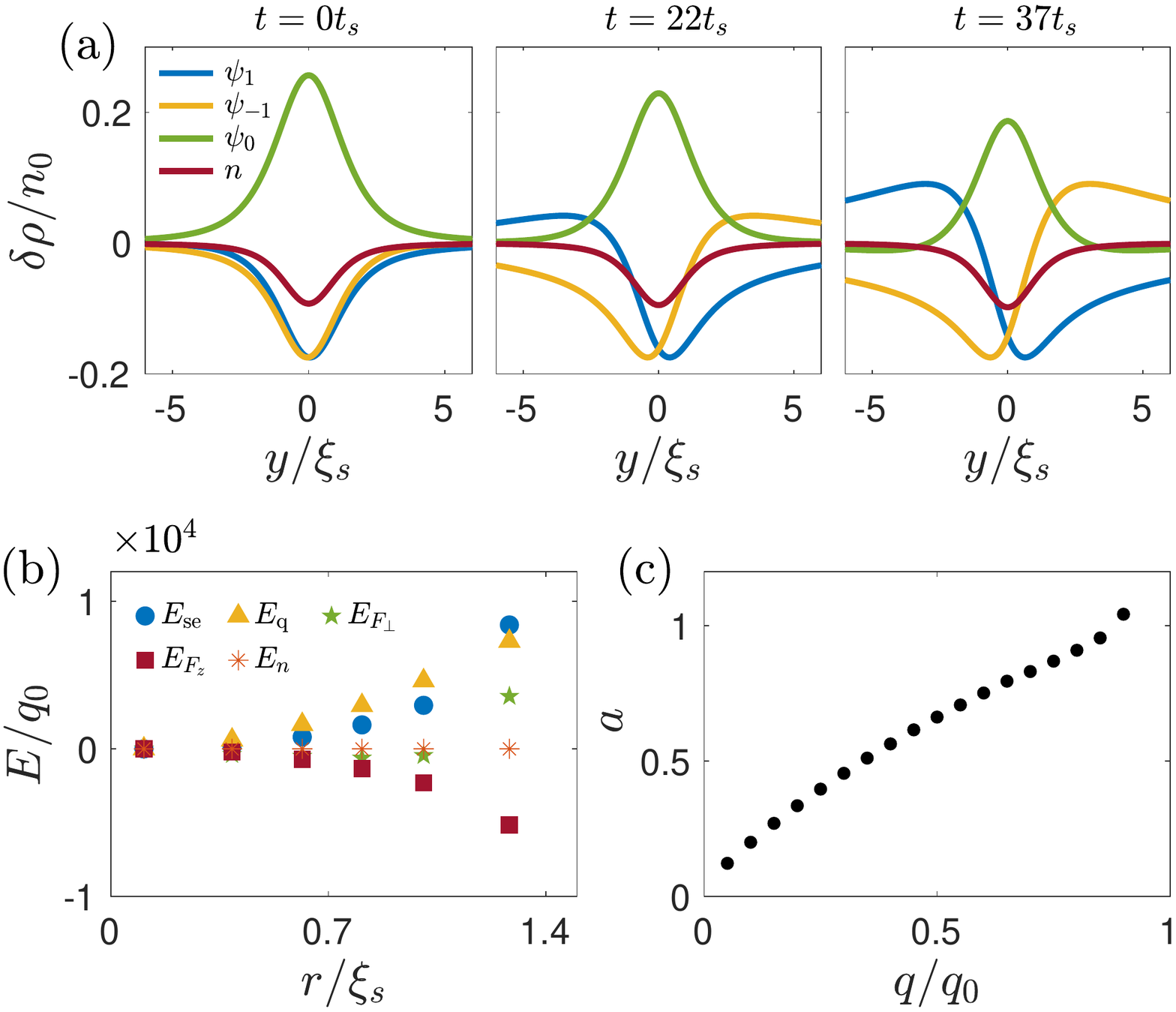}
\caption{\label{fig4}(a) The density profiles $\delta |\psi_m|^2=|\psi_m|^2-|\psi_m^\text{gs}|^2$ of the rightward moving (negatively charged) PCV core along the stretch axis. The profiles show that the stretching remains small compared to the PCV core (blue: $m=1$; yellow: $m=-1$; green: $m=0$). The total mass density $\delta n=n-n_0$ is also shown (red curve). (b) Within a $12\xi_s\times 12\xi_s$ area around either PCV, both the spin exchange energy $E_\text{se}$ (blue dots) and the quadratic Zeeman energy  $E_\text{q}$ (yellow triangles) increase as the PCV stretches. The increase in $E_\text{se}$ is compensated by a decrease in the remaining terms in $E_{F_\perp}$, such that $E_{F_\perp}$ (green stars) is close to constant. The energy $E_{F_z}$ (red squares) decreases as the PCV stretches and the $F_z$ dipole forms. The energy $E_n$ (orange asterisks) remains very close to constant and hence excitations of the total mass are decoupled from the PCV dynamics. The zero point of each energy has been shifted so that the smallest $r$ value plotted is at zero energy. For (a),(b) $q=0.3q_0$. (c) The spring constant $a$ from Fig.~\ref{fig2}(a) increases as $a\approx q/q_0$.}
\end{figure}

In~\cite{Turner2009} it was proposed that the stretch energy~\eqref{Hintstr} arises from the spin exchange interaction $\propto g_s\int d^2\mathbf{x}\,\operatorname{Re}(\psi_0^2\psi_1^*\psi_{-1}^*)$, which favours phase coherence between $\psi_1$ and $\psi_{-1}^*$. The spin exchange energy will increase as a PCV stretches and the circulating $\psi_{\pm 1}$ phases separate. We expect this energy to dominate for sufficiently large stretching where the cores of the $\psi_{\pm 1}$ vortices do not overlap and the $|\psi_m|^2$ profiles are independent of the stretching. However, for smaller stretching, energy terms involving $|\psi_m|^2$ may also be important. In Fig.~\ref{fig4}(a) we plot the density profiles $\delta |\psi_m|^2=|\psi_m|^2-|\psi_m^\text{gs}|^2$ of the three spin components $\psi_m$ through the centre of the rightward moving (negatively charged) PCV along $y$ (the stretch axis). Here $|\psi_m^\text{gs}|^2$ are the PCV-free ground-state densities computed from Eq.~\eqref{EPgs}. Clearly the stretching remains small compared to the PCV core size and the $|\psi_m|^2$ profiles change as the PCV stretches. Hence there will be a coupling between the stretch coordinate and energy terms involving $|\psi_m|^2$. Note that as the cores in the $\psi_{\pm 1}$ components separate, each core is partially filled by the other $\psi_{\mp 1}$ component, resulting in the dipole of $F_z$ magnetization in Fig.~\ref{fig1}(b). The density $\delta n=n-n_0$ is also plotted in Fig.~\ref{fig4}(a).

In Fig.~\ref{fig4}(b) we plot different contributions to the condensate energy, computed across a $12\xi_s\times 12\xi_s$ area $S$ centred at either PCV, for $q=0.3q_0$. The spin interaction energy consists of terms $E_{F_z}=(g_s/2)\int_S d^2\mathbf{x}\,F_z(\mathbf{x})^2$ and $E_{F_\perp}=(g_s/2)\int_S d^2\mathbf{x}\,|F_\perp|^2$. The latter term contains the spin exchange energy $E_\text{se}=2g_s\int_S d^2\mathbf{x}\,\operatorname{Re}(\psi_0^2\psi_1^*\psi_{-1}^*)$. Changes in $|\psi_m|^2$ may also change the quadratic Zeeman energy $E_\text{q}=q\int d^2\mathbf{x}\,(|\psi_1|^2+|\psi_{-1}|^2)$. Both the quadratic Zeeman energy $E_\text{q}=q\int d^2\mathbf{x}\,(|\psi_1|^2+|\psi_{-1}|^2)$ and the spin exchange energy $E_\text{se}=2g_s\int_S d^2\mathbf{x}\,\operatorname{Re}(\psi_0^2\psi_1^*\psi_{-1}^*)$ increase, roughly in proportion, as the PCV stretches. Therefore it is not clear whether the stretch energy arises from $E_\text{q}$ or $E_\text{se}$. The energy $E_{F_z}=2g_s\int_S d^2\mathbf{x}\,F_z(\mathbf{x})^2$ becomes increasingly more negative as the PCV stretches, as expected due to the formation of the $F_z$ dipole. The energy $E_{F_\perp}$ remains approximately constant and hence the increase in $E_\text{se}$ must coincide with a decrease of the remaining terms in $E_{F_\perp}$. We also plot the density interaction term $E_n=(g_n/2)\int_S d^2\mathbf{x}\,n(\mathbf{x})^2$, which remains very close to constant. Hence although the mass density variation across the PCV core is not negligible, see Fig.~\ref{fig4}(a), the mass density modes do not couple strongly to the PCV dynamics, as expected since $g_n\gg |g_s|$.

Although our analysis of Fig.~\ref{fig4}(b) provides some clues into the origin of the stretch energy, in general it is difficult to identify what excitations are contributing to $\mathcal{H}_r$ and what excitations are contributing to the the loss of the PCV energy $\mathcal{H}$ [Fig.~\ref{fig3}(b),(c)]. Furthermore, unlike in Fig.~\ref{fig4}(b), for $q\gtrsim 0.5q_0$ we find no clear trend in $E_\text{q}$ and $E_\text{se}$ with increasing $r$. This may relate to the qualitative change in the instability of a uniform polar state at $q=0.5q_0$~\cite{Saito2007a,uhlmann2007,lamacraft2007}. A more detailed analysis of the PCV core excitation spectrum, as has been done recently for nematic vortices in an antiferromagnetic spin-1 condensate~\cite{underwood2020}, may help elucidate more clearly what modes contribute to $\mathcal{H}_r$ and what modes contribute to the loss of $\mathcal{H}$. The stretch energy spring constant $a$, obtained from the fits to the results in Fig.~\ref{fig2}(a), increases close to linearly with increasing $q$, see Fig.~\ref{fig4}(c), with $a\approx q/q_0$ \footnote{The effective mass, not the bare mass, was fitted to simulation results in~\cite{williamson2016c}, giving a value of $a$ smaller than that in Fig.~\ref{fig4}(c).}. It is difficult, however, to associate this behaviour with $E_\text{q}$ or $E_\text{se}$, as the density and phase profiles within the PCV core are not uniform. A detailed study of a stretched PCV excitation spectrum may reveal modes that increase in energy as the PCV stretches and that scale with $q$ as in Fig.~\ref{fig4}(c), which would provide more insight into the basis of the PCV stretch energy.

\section{PCV annihilation}\label{annihilation}

\begin{figure}
\includegraphics[trim=0cm 4cm 0cm 4cm,clip=true,width=0.48\textwidth]{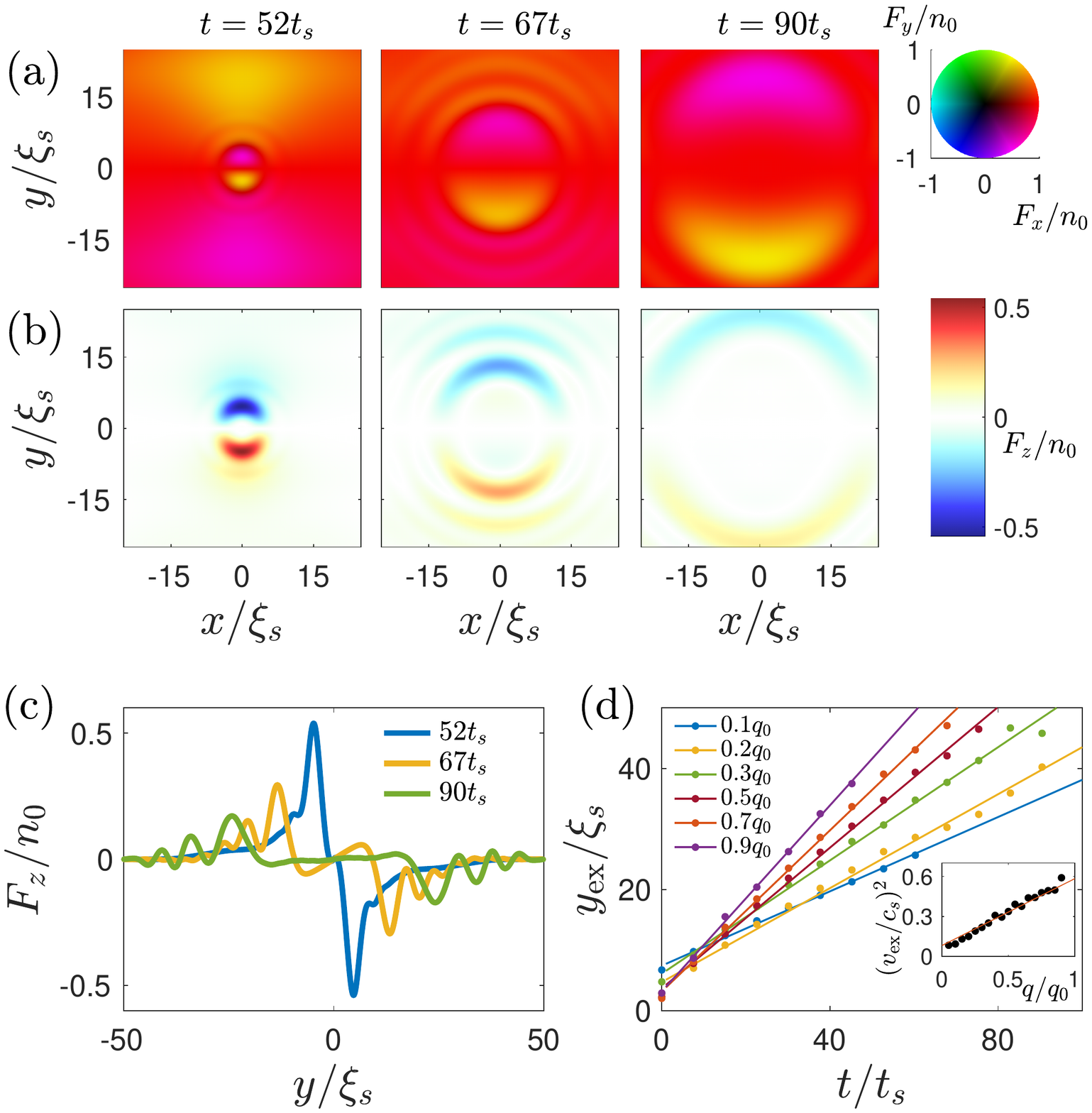}
\caption{\label{fig5}(a) Two colliding PCVs of opposite sign produce a transverse spin excitation that propagates outward from the annihilation point at the origin. The frames continue those displayed in Fig.~\ref{fig1}(a). (b) The corresponding $F_z$ excitation. (c) The $F_z(\mathbf{x})$ magnetization along the line $\mathbf{x}=(0,y)$, extracted from (b). The initial peak disperses as the excitation propagates [$t=52t_s$ (blue curve), $t=67t_s$ (yellow curve), $t=90t_s$ (green curve)]. For (a),(b),(c) $q=0.3q_0$. (d) The position $y_\text{ex}$ of the minimum of $F_z(0,y)$ increases linearly with time for different $q$ values (colored dots) [$q/q_0=(0.1,0.2,0.3,0.5,0.7,0.9)$ (blue, yellow, green, red, orange, purple)]. The excitation speed $v_\text{ex}$ can be extracted from a linear fit to the data points (colored solid lines). Inset: the speed $v_\text{ex}$ (black dots) follows the speed $v_\text{Bog}$ of long-wavelength Bogoliubov excitations of $F_z$ and $\operatorname{phase}(F_\perp)$ (red solid line).}
\end{figure}

Two oppositely charged PCVs attract until they collide and annihilate. In this section we examine the dynamics of the excitation produced following PCV annihilation. Figure~\ref{fig5} shows the evolving (a) transverse spin and (b) $F_z$ spin density, following the PCV annihilation imminent in Fig.~\ref{fig1}(a). The ring-shaped excitation that propagates out from the PCV annihilation point consists of variation in both the transverse spin direction and the $F_z$ spin density.

In Fig~\ref{fig5}(c) we plot $F_z(\mathbf{x})$ along the line $\mathbf{x}=(0,y)$, extracted from Fig.~\ref{fig5}(b). The excitation is initially dominated by a single peak, but this disperses as the excitation propagates out. We estimate the speed of the propagating waves from the propagation speed of the minimum of $F_z(0,y)$. The position $y_\text{ex}$ of this minimum travels at a constant speed $v_\text{ex}$, see Fig.~\ref{fig5}(d), which increases for increasing $q$. The speed is closely matched by the speed of long-wavelength Bogoliubov excitations of $F_z$, which dynamically couple to quadratic excitations of $\operatorname{phase}(F_\perp)$, see inset to Fig.~\ref{fig5}(d). For $g_n\gg |g_s|$ these excitations both have the dispersion relation $\hbar\omega_k=(q_0\xi_sk/2)\sqrt{\xi_s^2k^2+2q/q_0}$~\cite{barnett2011}, which travel at a speed $v_\text{Bog}=\omega_k/k$. Fitting $v_\text{Bog}$ to $v_\text{ex}$, we extract a small but non-zero wavevector $k\approx 2\pi\times 0.1\xi_s^{-1}$, which shifts the speed slightly from the sound ($k=0$) speed. In~\cite{williamson2019}, which studied the phase ordering dynamics following a quench to the easy-plane phase, it was found that out-of-equilibrium $F_z$ and $\operatorname{phase}(F_\perp)$ excitations remain after all PCVs have annihilated, leading to anomalously slow thermalization. However the origin of the excitations was not identified. Following the results from Fig.~\ref{fig5}, we expect that the excitations are produced from the PCV annihilation events during the PCV-driven phase ordering.

\section{PCV dynamics with non-zero axial magnetization}\label{axial}

\subsection{The damped Turner model with axial magnetization}
For $0<|p|<q$ in the BA phase, the ground state~\eqref{EPgs} is modified to~\cite{murata2007}
\begin{equation}\label{EPgsFz}
\Psi=e^{i\theta}e^{if_z\varphi}\sqrt{n_0}\left(\begin{array}{c}\cos\alpha\sin\beta\\\cos\beta \\\sin\alpha\sin\beta\end{array}\right),
\end{equation}
with
\begin{equation}
\begin{split}
\cos\beta=&\sqrt{\frac{(q^2-p^2)(p^2+q^2+qq_0)}{2q^3q_0}},\\
\cos 2\alpha=&\frac{2qp}{q^2+p^2}.
\end{split}
\end{equation}
In addition to the transverse magnetization $\mathbf{F}_\perp$, the state~\eqref{EPgsFz} has a non-zero axial magnetization,
\begin{equation}
F_z=\frac{n_0p(p^2-q^2+qq_0)}{q^2q_0}.
\end{equation}
A phase winding of the transverse spin gives rise to a PCV, with the state~\eqref{ansatz2} modified to
\begin{equation}\label{PCVz}
\Psi({\mathbf{x}},t)\approx \sqrt{n_0}\left(\begin{array}{c}\sin\beta\cos\alpha\prod_k g_{1k}(\mathbf{x})e^{-i\kappa_k \phi(\mathbf{x}-\mathbf{X}_{1k})}\\\prod_k\cos\beta g_{0k}(\mathbf{x})\\\sin\beta\sin\alpha\prod_k g_{-1k}(\mathbf{x})e^{i\kappa_k \phi(\mathbf{x}-\mathbf{X}_{-1k})}\end{array}\right).
\end{equation}
Carrying out an identical procedure that leads to the derivation of Eq.~\eqref{TMi} (see~\cite{williamson2016c}), we obtain
\begin{equation}\label{TMzi}
\begin{split}
\hat{\mathbf{z}}\times\dot{\mathbf{X}}_{mk}=&\sum_{j\ne k}\frac{m\hbar \kappa_j}{M}\frac{\mathbf{X}_{mk}-\mathbf{X}_{mj}}{|\mathbf{X}_{mk}-\mathbf{X}_{mj}|^2}\\
&-\frac{m\hbar\kappa_k}{2M[1+m\cos 2\alpha]}\nabla_{\mathbf{X}_{mk}}u(|\mathbf{r}_k|).
\end{split}
\end{equation}

According to Eqs.~\eqref{TMzi}, the Coulomb potential between two oppositely charged PCVs initially separated along the $x$-axis causes the $\psi_{\pm 1}$ components to stretch along the $y$-axis, like in the $F_z=0$ case. With non-zero $F_z$ magnetization, however, the prefactor of the balancing stretch force $\propto \nabla_{\mathbf{X}_{mk}}u(|\mathbf{r}_k|)$ is different for each of the $\psi_{\pm 1}$ components. As a result, the centres of circulation of the $\psi_{\pm 1}$ components move along the $x$-axis at different velocities. This results in an additional stretching along the $x$-axis, which needs to be balanced by a Magnus force along the $y$-axis. Hence in addition to the motion along the $x$-axis, the PCVs also move along the $y$-axis.

Continuing as for Eqs.~\eqref{TM}, we rewrite Eqs.~\eqref{TMzi} in terms of the coordinates $\mathbf{R}_k$ and $\mathbf{r}_k$. Including a damping term discussed in Sec.~\ref{dampedTurner} gives
\begin{subequations}\label{TMz}
\begin{align}
&\hat{\mathbf{z}}\times\dot{\mathbf{R}}_k=-\nabla_{\mathbf{r}_k}\frac{\hbar\kappa_k}{2M\sin^2 2\alpha}u(|\mathbf{r}_k|)\label{TMz1}\\
&\hat{\mathbf{z}}\times\left((1+\Gamma)\dot{\mathbf{r}}_k+2\cos 2\alpha\dot{\mathbf{R}}_k\right)=\frac{2\hbar \kappa_j}{M}\frac{\mathbf{R}_k-\mathbf{R}_j}{|\mathbf{R}_k-\mathbf{R}_j|^2}\label{TMz2}.
\end{align}
\end{subequations}

\subsection{PCV dynamics with small and large axial magnetization}

\begin{figure}
\includegraphics[trim=0cm 7cm 0cm 7cm,clip=true,width=0.48\textwidth]{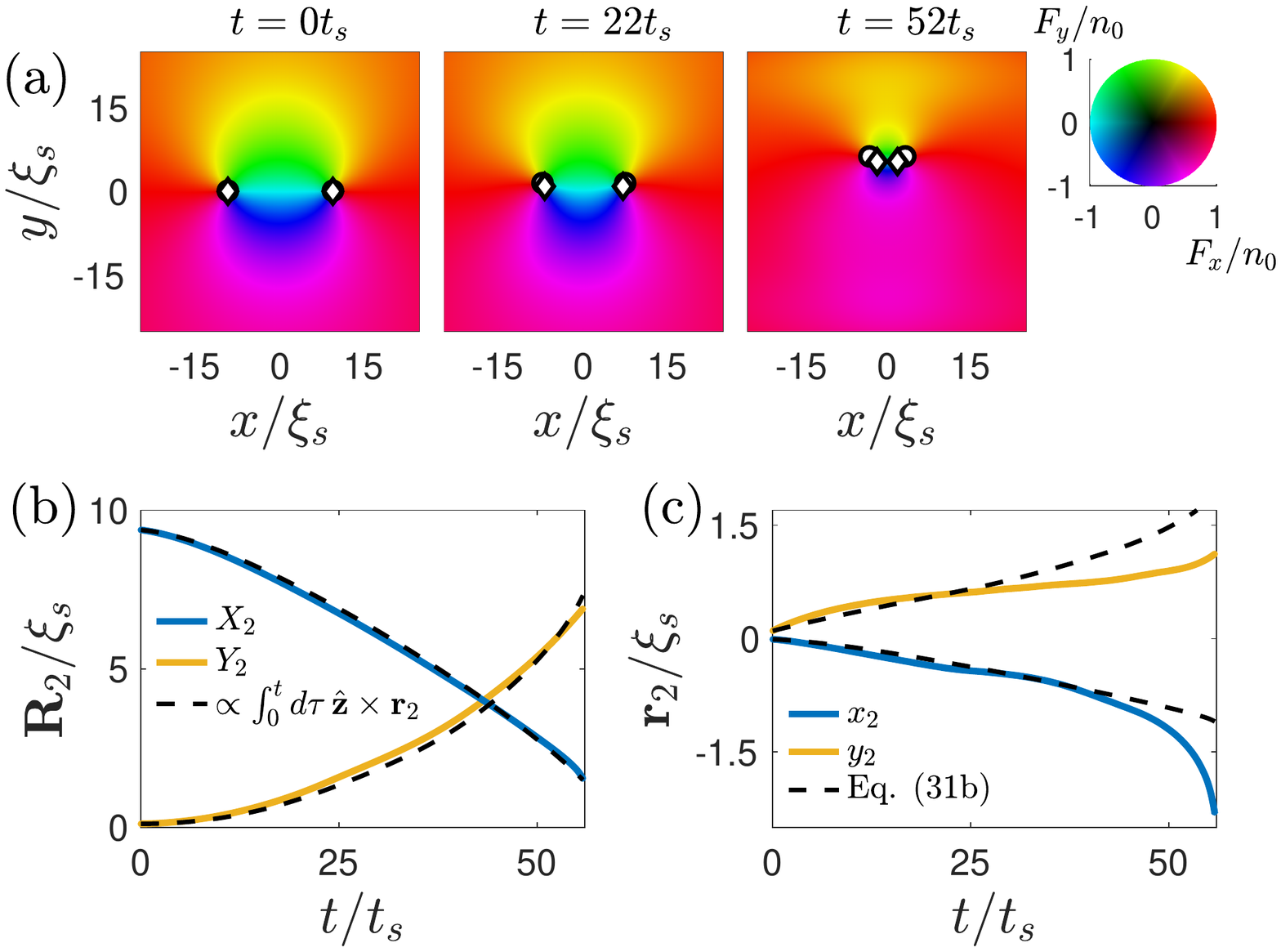}
\caption{\label{fig6}(a) A small nonzero $\langle F_z\rangle$ results in the two oppositely charged PCVs moving perpendicular to their separation, in addition to attracting. Circles(diamonds) mark the centre of circulation of the vortices in the $\psi_1(\psi_{-1})$ components. (b) The dynamics of the $X_2$ and $Y_2$ coordinates of $\mathbf{R}_2$ (colord solid curves) are proportional to $\int_0^td\tau\, \hat{\mathbf{z}}\times\mathbf{r}_2(\tau)$ (matching black dashed curves). (c) The stretch coordinates $\mathbf{r}_2=(x_2,y_2)$ (colord solid curves) are well described by Eq.~\eqref{TMz2} with damping rate $\Gamma\approx 3$ (matching black dashed lines). Results are for $\langle F_z\rangle=0.1n_0$, $q=0.3q_0$.}
\end{figure}

Figure~\ref{fig6} shows the PCV dynamics from GPE simulations as in Sec.~\ref{Turner}, but with the initial state~\eqref{PCVi} modified according to Eq.~\eqref{PCVz} such that $\langle F_z\rangle=0.1n_0$. Here $\langle F_z\rangle=A^{-1}\int d^2\mathbf{x}\,F_z$ is the spatially averaged $F_z$ density ($A$ is the system area). As predicted, the PCVs not only attract but also move transverse to their separation, see Fig.~\ref{fig6}(a). The dynamics of $\mathbf{R}_k$ follows closely the integral $[(\hbar\kappa_k)/(2M\sin^2 2\alpha)]\int_0^td\tau\, \hat{\mathbf{z}}\times\nabla_{\mathbf{r}_k}u(|\mathbf{r}_k(\tau)|)$ using the potential~\eqref{uapprox}, see Fig.~\ref{fig6}(b). The spring constant $a$ is fitted using GPE values for $|\mathbf{R}_k|$ and $|\mathbf{r}_k|$ (note $\hat{\mathbf{z}}\times\nabla_{\mathbf{r}_k}r_k^2=2\hat{\mathbf{z}}\times\mathbf{r}_k$). The dynamics of $\mathbf{r}_2$ is shown in Fig.~\ref{fig6}(c), compared with the prediction of Eq.~\eqref{TMz2} using simulation results for $\mathbf{R}_2$. Equation~\eqref{TMz2} describes the dynamics of $\mathbf{r}_2$ reasonably well with damping rate $\Gamma\approx 3$.

\begin{figure}
\includegraphics[trim=0cm 7cm 0cm 7cm,clip=true,width=0.48\textwidth]{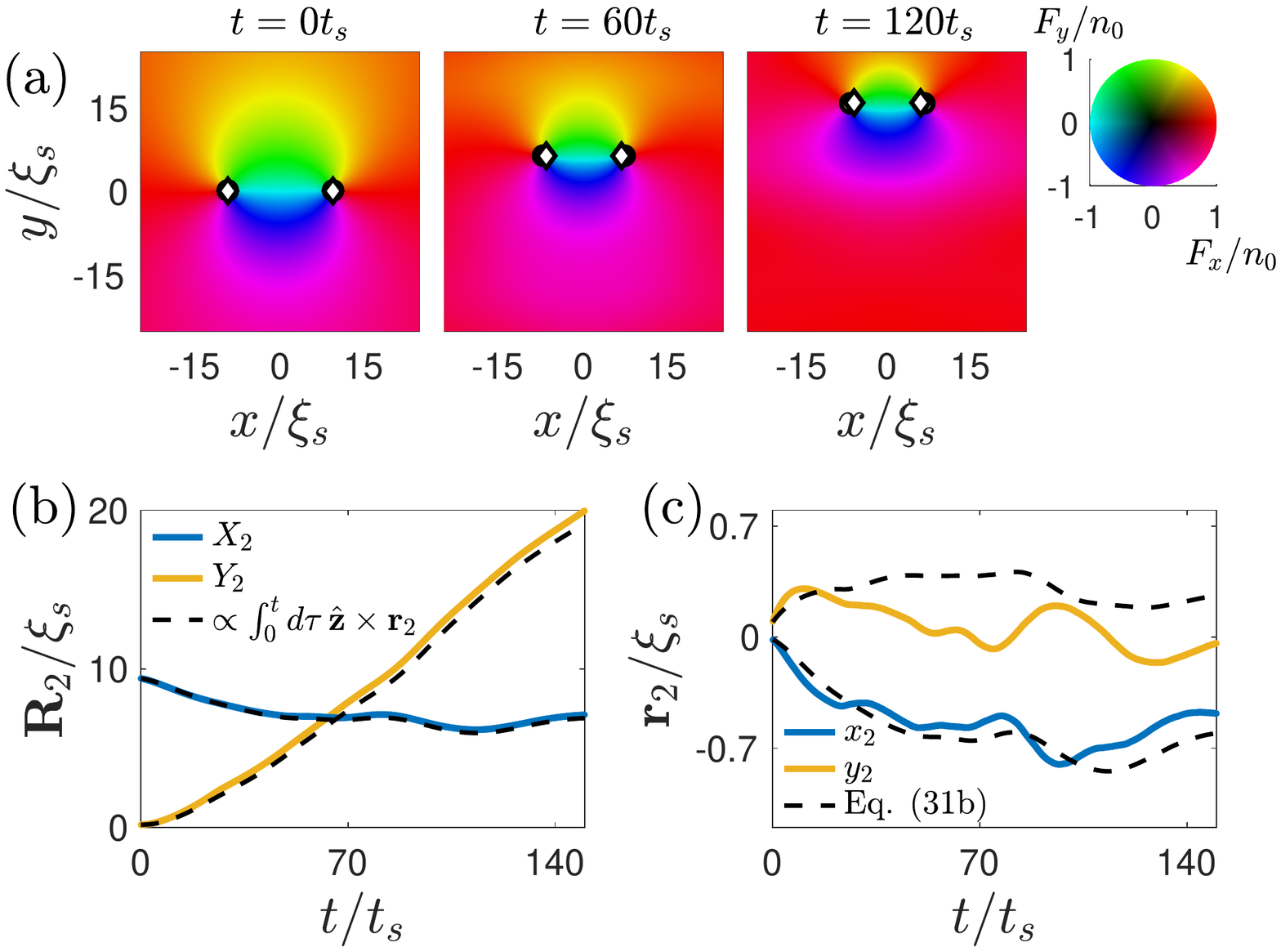}
\caption{\label{fig7}(a) For larger $\langle F_z\rangle$ magnetization, $\mathbf{r}_k$ rotates until it is parallel with $\mathbf{R}_k$ and the PCVs move perpendicular to their separation, like scalar vortices. Circles(diamonds) mark the centre of circulation of the vortices in the $\psi_1(\psi_{-1})$ components. (b) The dynamics of the $X_2$ and $Y_2$ coordinates of $\mathbf{R}_2$ (colord solid curves) are proportional to $\int_0^td\tau\, \hat{\mathbf{z}}\times\mathbf{r}_2(\tau)$ (matching black dashed curves), with $\dot{X}_2\approx 0$ and constant $\dot{Y}_2$. (c) The stretch $\mathbf{r}_2$ (colored solid curves) reaches an approximate steady-state with $y_2\approx 0$, giving rise to the scalar vortex behaviour observed in (a),(b). The results are qualitatively described by Eq.~\eqref{TMz2} (matching black dashed lines) with damping rate $\Gamma\approx 3$. Results are for $\langle F_z\rangle=0.2n_0$, $q=0.3q_0$.}
\end{figure}

Figure~\ref{fig7} shows the PCV dynamics from GPE simulations with $\langle F_z\rangle=0.2n_0$. For this higher magnetization, we find that the stretch coordinate $\mathbf{r}_k$ rotates until it is parallel $\mathbf{R}_k$, see Fig.~\ref{fig7}(a). Hence the attractive force between the PCVs becomes negligible and instead the two PCVs move parallel, analogous to scalar vortices. (This effect becomes more pronounced as $|\langle F_z\rangle |$ is increased.) The dynamics of $\mathbf{R}_k$ follows closely the integral $[(\hbar\kappa_k)/(2M\sin^2 2\alpha)]\int_0^td\tau\, \hat{\mathbf{z}}\times\nabla_{\mathbf{r}_k}u(|\mathbf{r}_k(\tau)|)$ using the potential~\eqref{uapprox}, see Fig.~\ref{fig7}(b). We find that $a/\sin^2 (2\beta)$ has little dependence on $\langle F_z\rangle$ for $|\langle F_z\rangle|/n_0\lesssim 0.2$, with $a/\sin^2(2\beta)=0.46,0.45,0.48$ for $\langle F_z\rangle /n_0=0,0.1,0.2$ respectively. As observed in Fig.~\ref{fig7}(a), $\mathbf{r}_2$ rotates until $y_2\approx 0$, see Fig.~\ref{fig7}(c). The small stretch $x_2$ also approximates a steady state. For this larger magnetization, the prediction of Eq.~\eqref{TMz2}, using simulation results for $\mathbf{R}_2$, is less accurate, but still provides a qualitative description of the dynamics with damping rate $\Gamma\approx 3$. The precise nature of the oscillations in Fig.~\ref{fig7}(c) are sensitive to the length of the initial damped evolution, but the overall trend of the curves remains unchanged.

The crossover to scalar vortex dynamics is interesting and warrants further discussion. Setting $\dot{\mathbf{r}}_k=0$ in Eq.~\eqref{TMz2} gives,
\begin{equation}\label{TMzs}
\hat{\mathbf{z}}\times\dot{\mathbf{R}}_k=\sum_{j\ne k}\frac{\hbar \kappa_j}{M\cos 2\alpha}\frac{\mathbf{R}_k-\mathbf{R}_j}{|\mathbf{R}_k-\mathbf{R}_j|^2},
\end{equation}
identical in form to the point-vortex equations for scalar vortices, Eq.~\eqref{scalarvortex}. For the two-PCV setup in Fig.~\ref{fig6},~\ref{fig7}, Eq.~\eqref{TMzs} allows for constant $\dot{\mathbf{R}}_k$ and hence a constant $\mathbf{r}_k$ is compatible with Eq.~\eqref{TMz1}. We expect that this crossover from PCV dynamics to scalar vortex dynamics will have interesting consequences in PCV-driven phase ordering, following quenches to the BA phase. For zero $\langle F_z\rangle$, the phase ordering is governed by a dynamic critical exponent $z=1$~\cite{williamson2016a,williamson2016b}. For sufficiently large $|\langle F_z\rangle |$, we expect the attractive force between PCVs to become negligible compared to scalar-vortex damping~\cite{billam2015}, and the PCVs will behave like scalar vortices. In this \emph{scalar-vortex regime} we expect critical behaviour that follows that of a scalar condensate, with $z=2$~\cite{karl2017,comaron2019,groszek2020}.

Considering an arbitrary $|\langle F_z\rangle |>0$, we envisage two possible phase-ordering scenarios, dependent on whether or not the scalar-vortex regime is reached prior to PCV annihilation. Scenario 1: The scalar vortex regime is \emph{always} reached prior to PCV annihilation, given a sufficiently large PCV separation, and hence the long-time limit of phase ordering with nonzero $\langle F_z\rangle$ would be that of a scalar system. This is analogous to the small-$q$ phase ordering around the isotropic ferromagnetic phase~\cite{williamson2017}. Scenario 2: There exists some critical value of $|\langle F_z\rangle |$ below which the scalar-vortex regime is \emph{not} reached prior to PCV annihilation, for any initial PCV separation, and hence the BA phase would be divided into two dynamic universality classes either side of this critical value of $|\langle F_z\rangle |$. A many-PCV simulation that allows for sound-induced damping~\cite{billam2015} would be needed to definitively determine which scenario occurs, however identifying regimes where Eq.~\eqref{TMzs} is reached from Eq.~\eqref{TMz} may give insight into this question. For example, for the two-PCV setup in Figs.~\ref{fig6},~\ref{fig7}, one could compare $t_\text{coll}$ with the time for $\mathbf{r}_k$ to rotate parallel to $\mathbf{R}_k$, as a function of $\langle F_z\rangle$ and initial PCV separation.

\section{Summary and outlook}\label{conclusion}
In this work we have systematically studied the accuracy of the Turner point-vortex model at describing the dynamics of two oppositely charged PCVs, via comparisons with GPE simulations. While the bare Turner model, as introduced in~\cite{Turner2009}, shows discrepancies with GPE results, the agreement can be vastly improved by introducing a phenomenological damping of the PCV stretch coordinate. The damping arises from coupling to other excitations not accounted for in the point-vortex phase space, and reduces the PCV stretching. The net effect on the mesoscopic dynamics, however, is to simply increase the PCVs bare mass to give an effective mass that increases the PCV collision time. In addition, we identified the excitations produced following PCV annihilation and extended the Turner model to describe PCV dynamics with axial magnetization, both of which have applications to the phase ordering of spin-1 BECs.

Some important questions raised by this work remain to be answered. Firstly, can the modes that contribute to the PCV stretch energy be identified; and secondly, can the modes that are responsible for the damping of the stretch coordinate be identified. In addition, a derivation of the Turner point-vortex model from the hydrodynamic formulation of a ferromagnetic spin-1 condensate~\cite{Yukawa2012} may offer additional insights into the fluid dynamics of the Turner model, analogous to the derivation of the scalar point-vortex model from the Euler equation~\cite{geldhauser2019}. For example, the scalar point-vortex model, which describes mass currents, reflects conservation of momentum. As the circulating transverse spin around a PCV corresponds to an $F_z$ magnetization current, Eqs.~\eqref{TM} and their modifications Eqs.~\eqref{TMz} may arise from an analogous conservation law for $F_z$ ``spin momentum''.

A useful property of PCVs is that they can be observed \emph{in situ} in experiments~\cite{Sadler2006a} using non-destructive phase contrast imaging~\cite{higbie2005}. To observe the attraction of two PCVs, the axial magnetization should be kept small, $|\langle F_z\rangle |\lesssim 0.2n_0$. The effective mass of two PCVs could be obtained from measuring the PCV collision time. The spring constant $a$ could be measured with highly resolved absorption imaging~\cite{muessel2013,prufer2018} of the $F_z$ dipole, which forms as the PCV stretches. Combined with the effective mass, this would allow extraction of the damping rate $\Gamma$.

A host of possible applications of PCVs could also be explored, for example few-body PCV dynamics, PCV driven turbulence, emergent hydrodynamics~\cite{wiegmann2014,xiaoquan2017,stockdale2020}, and thermodynamic properties of PCV systems. Also, the coupling to other dynamical modes through the PCV stretching and the associated modification of the PCV mass raises the question of whether PCV dynamics could be manipulated via the use of additional interactions or external fields. The success of our damped Turner model provides a useful starting point for these explorations.

\section*{Acknowledgements}
We acknowledge support from the Marsden Fund of the Royal Society of New Zealand.


\begin{thebibliography}{97}%
\makeatletter
\providecommand \@ifxundefined [1]{%
 \@ifx{#1\undefined}
}%
\providecommand \@ifnum [1]{%
 \ifnum #1\expandafter \@firstoftwo
 \else \expandafter \@secondoftwo
 \fi
}%
\providecommand \@ifx [1]{%
 \ifx #1\expandafter \@firstoftwo
 \else \expandafter \@secondoftwo
 \fi
}%
\providecommand \natexlab [1]{#1}%
\providecommand \enquote  [1]{``#1''}%
\providecommand \bibnamefont  [1]{#1}%
\providecommand \bibfnamefont [1]{#1}%
\providecommand \citenamefont [1]{#1}%
\providecommand \href@noop [0]{\@secondoftwo}%
\providecommand \href [0]{\begingroup \@sanitize@url \@href}%
\providecommand \@href[1]{\@@startlink{#1}\@@href}%
\providecommand \@@href[1]{\endgroup#1\@@endlink}%
\providecommand \@sanitize@url [0]{\catcode `\\12\catcode `\$12\catcode
  `\&12\catcode `\#12\catcode `\^12\catcode `\_12\catcode `\%12\relax}%
\providecommand \@@startlink[1]{}%
\providecommand \@@endlink[0]{}%
\providecommand \url  [0]{\begingroup\@sanitize@url \@url }%
\providecommand \@url [1]{\endgroup\@href {#1}{\urlprefix }}%
\providecommand \urlprefix  [0]{URL }%
\providecommand \Eprint [0]{\href }%
\providecommand \doibase [0]{http://dx.doi.org/}%
\providecommand \selectlanguage [0]{\@gobble}%
\providecommand \bibinfo  [0]{\@secondoftwo}%
\providecommand \bibfield  [0]{\@secondoftwo}%
\providecommand \translation [1]{[#1]}%
\providecommand \BibitemOpen [0]{}%
\providecommand \bibitemStop [0]{}%
\providecommand \bibitemNoStop [0]{.\EOS\space}%
\providecommand \EOS [0]{\spacefactor3000\relax}%
\providecommand \BibitemShut  [1]{\csname bibitem#1\endcsname}%
\let\auto@bib@innerbib\@empty
\bibitem [{\citenamefont {Kawaguchi}\ and\ \citenamefont
  {Ueda}(2012)}]{Kawaguchi2012R}%
  \BibitemOpen
  \bibfield  {author} {\bibinfo {author} {\bibfnamefont {Y.}~\bibnamefont
  {Kawaguchi}}\ and\ \bibinfo {author} {\bibfnamefont {M.}~\bibnamefont
  {Ueda}},\ }\href {\doibase 10.1016/j.physrep.2012.07.005} {\bibfield
  {journal} {\bibinfo  {journal} {Physics Reports}\ }\textbf {\bibinfo {volume}
  {520}},\ \bibinfo {pages} {253 } (\bibinfo {year} {2012})}\BibitemShut
  {NoStop}%
\bibitem [{\citenamefont {Stamper-Kurn}\ and\ \citenamefont
  {Ueda}(2013)}]{StamperKurn2013a}%
  \BibitemOpen
  \bibfield  {author} {\bibinfo {author} {\bibfnamefont {D.~M.}\ \bibnamefont
  {Stamper-Kurn}}\ and\ \bibinfo {author} {\bibfnamefont {M.}~\bibnamefont
  {Ueda}},\ }\href {\doibase 10.1103/RevModPhys.85.1191} {\bibfield  {journal}
  {\bibinfo  {journal} {Rev. Mod. Phys.}\ }\textbf {\bibinfo {volume} {85}},\
  \bibinfo {pages} {1191} (\bibinfo {year} {2013})}\BibitemShut {NoStop}%
\bibitem [{\citenamefont {Kibble}(1976)}]{Kibble1976a}%
  \BibitemOpen
  \bibfield  {author} {\bibinfo {author} {\bibfnamefont {T.~W.~B.}\
  \bibnamefont {Kibble}},\ }\href {http://stacks.iop.org/0305-4470/9/i=8/a=029}
  {\bibfield  {journal} {\bibinfo  {journal} {J. Phys. A: Math. Gen.}\ }\textbf
  {\bibinfo {volume} {9}},\ \bibinfo {pages} {1387} (\bibinfo {year}
  {1976})}\BibitemShut {NoStop}%
\bibitem [{\citenamefont {Zurek}(1985)}]{Zurek1985a}%
  \BibitemOpen
  \bibfield  {author} {\bibinfo {author} {\bibfnamefont {W.~H.}\ \bibnamefont
  {Zurek}},\ }\href@noop {} {\bibfield  {journal} {\bibinfo  {journal}
  {Nature}\ }\textbf {\bibinfo {volume} {317}},\ \bibinfo {pages} {505}
  (\bibinfo {year} {1985})}\BibitemShut {NoStop}%
\bibitem [{\citenamefont {Bray}(1994)}]{Bray1994}%
  \BibitemOpen
  \bibfield  {author} {\bibinfo {author} {\bibfnamefont {A.}~\bibnamefont
  {Bray}},\ }\href {\doibase 10.1080/00018739400101505} {\bibfield  {journal}
  {\bibinfo  {journal} {Adv. Phys.}\ }\textbf {\bibinfo {volume} {43}},\
  \bibinfo {pages} {357} (\bibinfo {year} {1994})}\BibitemShut {NoStop}%
\bibitem [{\citenamefont {Bradley}\ and\ \citenamefont
  {Anderson}(2012)}]{bradley2012}%
  \BibitemOpen
  \bibfield  {author} {\bibinfo {author} {\bibfnamefont {A.~S.}\ \bibnamefont
  {Bradley}}\ and\ \bibinfo {author} {\bibfnamefont {B.~P.}\ \bibnamefont
  {Anderson}},\ }\href {\doibase 10.1103/PhysRevX.2.041001} {\bibfield
  {journal} {\bibinfo  {journal} {Phys. Rev. X}\ }\textbf {\bibinfo {volume}
  {2}},\ \bibinfo {pages} {041001} (\bibinfo {year} {2012})}\BibitemShut
  {NoStop}%
\bibitem [{\citenamefont {Barenghi}\ \emph {et~al.}(2001)\citenamefont
  {Barenghi}, \citenamefont {Donnelly},\ and\ \citenamefont
  {Vinen}}]{barenghi2001}%
  \BibitemOpen
  \bibfield  {author} {\bibinfo {author} {\bibfnamefont {C.}~\bibnamefont
  {Barenghi}}, \bibinfo {author} {\bibfnamefont {R.}~\bibnamefont {Donnelly}},
  \ and\ \bibinfo {author} {\bibfnamefont {W.}~\bibnamefont {Vinen}},\ }\href
  {\doibase 10.1007/3-540-45542-6} {\emph {\bibinfo {title} {Quantized Vortex
  Dynamics and Superfluid Turbulence}}}\ (\bibinfo  {publisher} {Springer},\
  \bibinfo {year} {2001})\BibitemShut {NoStop}%
\bibitem [{\citenamefont {Yip}(1999)}]{yip1999}%
  \BibitemOpen
  \bibfield  {author} {\bibinfo {author} {\bibfnamefont {S.-K.}\ \bibnamefont
  {Yip}},\ }\href {\doibase 10.1103/PhysRevLett.83.4677} {\bibfield  {journal}
  {\bibinfo  {journal} {Phys. Rev. Lett.}\ }\textbf {\bibinfo {volume} {83}},\
  \bibinfo {pages} {4677} (\bibinfo {year} {1999})}\BibitemShut {NoStop}%
\bibitem [{\citenamefont {Isoshima}\ \emph {et~al.}(2001)\citenamefont
  {Isoshima}, \citenamefont {Machida},\ and\ \citenamefont
  {Ohmi}}]{isoshima2001}%
  \BibitemOpen
  \bibfield  {author} {\bibinfo {author} {\bibfnamefont {T.}~\bibnamefont
  {Isoshima}}, \bibinfo {author} {\bibfnamefont {K.}~\bibnamefont {Machida}}, \
  and\ \bibinfo {author} {\bibfnamefont {T.}~\bibnamefont {Ohmi}},\ }\href
  {https://doi.org/10.1143/JPSJ.70.1604} {\bibfield  {journal} {\bibinfo
  {journal} {J. Phys. Soc. Jpn.}\ }\textbf {\bibinfo {volume} {70}},\ \bibinfo
  {pages} {1604} (\bibinfo {year} {2001})}\BibitemShut {NoStop}%
\bibitem [{\citenamefont {Stoof}\ \emph {et~al.}(2001)\citenamefont {Stoof},
  \citenamefont {Vliegen},\ and\ \citenamefont {Al~Khawaja}}]{stoof2001}%
  \BibitemOpen
  \bibfield  {author} {\bibinfo {author} {\bibfnamefont {H.~T.~C.}\
  \bibnamefont {Stoof}}, \bibinfo {author} {\bibfnamefont {E.}~\bibnamefont
  {Vliegen}}, \ and\ \bibinfo {author} {\bibfnamefont {U.}~\bibnamefont
  {Al~Khawaja}},\ }\href {\doibase 10.1103/PhysRevLett.87.120407} {\bibfield
  {journal} {\bibinfo  {journal} {Phys. Rev. Lett.}\ }\textbf {\bibinfo
  {volume} {87}},\ \bibinfo {pages} {120407} (\bibinfo {year}
  {2001})}\BibitemShut {NoStop}%
\bibitem [{\citenamefont {Al~Khawaja}\ and\ \citenamefont
  {Stoof}(2001)}]{khawaja2001a}%
  \BibitemOpen
  \bibfield  {author} {\bibinfo {author} {\bibfnamefont {U.}~\bibnamefont
  {Al~Khawaja}}\ and\ \bibinfo {author} {\bibfnamefont {H.}~\bibnamefont
  {Stoof}},\ }\href {https://doi.org/10.1038/35082010} {\bibfield  {journal}
  {\bibinfo  {journal} {Nature}\ }\textbf {\bibinfo {volume} {411}},\ \bibinfo
  {pages} {918} (\bibinfo {year} {2001})}\BibitemShut {NoStop}%
\bibitem [{\citenamefont {Khawaja}\ and\ \citenamefont
  {Stoof}(2001)}]{khawaja2001b}%
  \BibitemOpen
  \bibfield  {author} {\bibinfo {author} {\bibfnamefont {U.~A.}\ \bibnamefont
  {Khawaja}}\ and\ \bibinfo {author} {\bibfnamefont {H.~T.~C.}\ \bibnamefont
  {Stoof}},\ }\href {\doibase 10.1103/PhysRevA.64.043612} {\bibfield  {journal}
  {\bibinfo  {journal} {Phys. Rev. A}\ }\textbf {\bibinfo {volume} {64}},\
  \bibinfo {pages} {043612} (\bibinfo {year} {2001})}\BibitemShut {NoStop}%
\bibitem [{\citenamefont {Zhou}(2001)}]{zhou2001}%
  \BibitemOpen
  \bibfield  {author} {\bibinfo {author} {\bibfnamefont {F.}~\bibnamefont
  {Zhou}},\ }\href {\doibase 10.1103/PhysRevLett.87.080401} {\bibfield
  {journal} {\bibinfo  {journal} {Phys. Rev. Lett.}\ }\textbf {\bibinfo
  {volume} {87}},\ \bibinfo {pages} {080401} (\bibinfo {year}
  {2001})}\BibitemShut {NoStop}%
\bibitem [{\citenamefont {Mizushima}\ \emph {et~al.}(2002)\citenamefont
  {Mizushima}, \citenamefont {Machida},\ and\ \citenamefont
  {Kita}}]{mizushima2002}%
  \BibitemOpen
  \bibfield  {author} {\bibinfo {author} {\bibfnamefont {T.}~\bibnamefont
  {Mizushima}}, \bibinfo {author} {\bibfnamefont {K.}~\bibnamefont {Machida}},
  \ and\ \bibinfo {author} {\bibfnamefont {T.}~\bibnamefont {Kita}},\ }\href
  {\doibase 10.1103/PhysRevLett.89.030401} {\bibfield  {journal} {\bibinfo
  {journal} {Phys. Rev. Lett.}\ }\textbf {\bibinfo {volume} {89}},\ \bibinfo
  {pages} {030401} (\bibinfo {year} {2002})}\BibitemShut {NoStop}%
\bibitem [{\citenamefont {M{\"a}kel{\"a}}\ \emph {et~al.}(2003)\citenamefont
  {M{\"a}kel{\"a}}, \citenamefont {Zhang},\ and\ \citenamefont
  {Suominen}}]{makela2003}%
  \BibitemOpen
  \bibfield  {author} {\bibinfo {author} {\bibfnamefont {H.}~\bibnamefont
  {M{\"a}kel{\"a}}}, \bibinfo {author} {\bibfnamefont {Y.}~\bibnamefont
  {Zhang}}, \ and\ \bibinfo {author} {\bibfnamefont {K.-A.}\ \bibnamefont
  {Suominen}},\ }\href {https://doi.org/10.1088/0305-4470/36/32/302} {\bibfield
   {journal} {\bibinfo  {journal} {J. Phys. A: Math. Gen.}\ }\textbf {\bibinfo
  {volume} {36}},\ \bibinfo {pages} {8555} (\bibinfo {year}
  {2003})}\BibitemShut {NoStop}%
\bibitem [{\citenamefont {Ruostekoski}\ and\ \citenamefont
  {Anglin}(2003)}]{ruostekoski2003}%
  \BibitemOpen
  \bibfield  {author} {\bibinfo {author} {\bibfnamefont {J.}~\bibnamefont
  {Ruostekoski}}\ and\ \bibinfo {author} {\bibfnamefont {J.~R.}\ \bibnamefont
  {Anglin}},\ }\href {\doibase 10.1103/PhysRevLett.91.190402} {\bibfield
  {journal} {\bibinfo  {journal} {Phys. Rev. Lett.}\ }\textbf {\bibinfo
  {volume} {91}},\ \bibinfo {pages} {190402} (\bibinfo {year}
  {2003})}\BibitemShut {NoStop}%
\bibitem [{\citenamefont {Semenoff}\ and\ \citenamefont
  {Zhou}(2007)}]{semenoff2007}%
  \BibitemOpen
  \bibfield  {author} {\bibinfo {author} {\bibfnamefont {G.~W.}\ \bibnamefont
  {Semenoff}}\ and\ \bibinfo {author} {\bibfnamefont {F.}~\bibnamefont
  {Zhou}},\ }\href {\doibase 10.1103/PhysRevLett.98.100401} {\bibfield
  {journal} {\bibinfo  {journal} {Phys. Rev. Lett.}\ }\textbf {\bibinfo
  {volume} {98}},\ \bibinfo {pages} {100401} (\bibinfo {year}
  {2007})}\BibitemShut {NoStop}%
\bibitem [{\citenamefont {Barnett}\ \emph {et~al.}(2007)\citenamefont
  {Barnett}, \citenamefont {Turner},\ and\ \citenamefont
  {Demler}}]{barnett2007}%
  \BibitemOpen
  \bibfield  {author} {\bibinfo {author} {\bibfnamefont {R.}~\bibnamefont
  {Barnett}}, \bibinfo {author} {\bibfnamefont {A.}~\bibnamefont {Turner}}, \
  and\ \bibinfo {author} {\bibfnamefont {E.}~\bibnamefont {Demler}},\ }\href
  {\doibase 10.1103/PhysRevA.76.013605} {\bibfield  {journal} {\bibinfo
  {journal} {Phys. Rev. A}\ }\textbf {\bibinfo {volume} {76}},\ \bibinfo
  {pages} {013605} (\bibinfo {year} {2007})}\BibitemShut {NoStop}%
\bibitem [{\citenamefont {Kawaguchi}\ \emph {et~al.}(2008)\citenamefont
  {Kawaguchi}, \citenamefont {Nitta},\ and\ \citenamefont
  {Ueda}}]{kawaguchi2008}%
  \BibitemOpen
  \bibfield  {author} {\bibinfo {author} {\bibfnamefont {Y.}~\bibnamefont
  {Kawaguchi}}, \bibinfo {author} {\bibfnamefont {M.}~\bibnamefont {Nitta}}, \
  and\ \bibinfo {author} {\bibfnamefont {M.}~\bibnamefont {Ueda}},\ }\href
  {\doibase 10.1103/PhysRevLett.100.180403} {\bibfield  {journal} {\bibinfo
  {journal} {Phys. Rev. Lett.}\ }\textbf {\bibinfo {volume} {100}},\ \bibinfo
  {pages} {180403} (\bibinfo {year} {2008})}\BibitemShut {NoStop}%
\bibitem [{\citenamefont {Huhtam\"aki}\ \emph {et~al.}(2009)\citenamefont
  {Huhtam\"aki}, \citenamefont {Simula}, \citenamefont {Kobayashi},\ and\
  \citenamefont {Machida}}]{huhtamaki2009}%
  \BibitemOpen
  \bibfield  {author} {\bibinfo {author} {\bibfnamefont {J.~A.~M.}\
  \bibnamefont {Huhtam\"aki}}, \bibinfo {author} {\bibfnamefont {T.~P.}\
  \bibnamefont {Simula}}, \bibinfo {author} {\bibfnamefont {M.}~\bibnamefont
  {Kobayashi}}, \ and\ \bibinfo {author} {\bibfnamefont {K.}~\bibnamefont
  {Machida}},\ }\href {\doibase 10.1103/PhysRevA.80.051601} {\bibfield
  {journal} {\bibinfo  {journal} {Phys. Rev. A}\ }\textbf {\bibinfo {volume}
  {80}},\ \bibinfo {pages} {051601} (\bibinfo {year} {2009})}\BibitemShut
  {NoStop}%
\bibitem [{\citenamefont {Lovegrove}\ \emph {et~al.}(2014)\citenamefont
  {Lovegrove}, \citenamefont {Borgh},\ and\ \citenamefont
  {Ruostekoski}}]{lovegrove2014}%
  \BibitemOpen
  \bibfield  {author} {\bibinfo {author} {\bibfnamefont {J.}~\bibnamefont
  {Lovegrove}}, \bibinfo {author} {\bibfnamefont {M.~O.}\ \bibnamefont
  {Borgh}}, \ and\ \bibinfo {author} {\bibfnamefont {J.}~\bibnamefont
  {Ruostekoski}},\ }\href {\doibase 10.1103/PhysRevLett.112.075301} {\bibfield
  {journal} {\bibinfo  {journal} {Phys. Rev. Lett.}\ }\textbf {\bibinfo
  {volume} {112}},\ \bibinfo {pages} {075301} (\bibinfo {year}
  {2014})}\BibitemShut {NoStop}%
\bibitem [{\citenamefont {Borgh}\ and\ \citenamefont
  {Ruostekoski}(2016)}]{borgh2016}%
  \BibitemOpen
  \bibfield  {author} {\bibinfo {author} {\bibfnamefont {M.~O.}\ \bibnamefont
  {Borgh}}\ and\ \bibinfo {author} {\bibfnamefont {J.}~\bibnamefont
  {Ruostekoski}},\ }\href {\doibase 10.1103/PhysRevLett.117.275302} {\bibfield
  {journal} {\bibinfo  {journal} {Phys. Rev. Lett.}\ }\textbf {\bibinfo
  {volume} {117}},\ \bibinfo {pages} {275302} (\bibinfo {year}
  {2016})}\BibitemShut {NoStop}%
\bibitem [{\citenamefont {Tiurev}\ \emph {et~al.}(2018)\citenamefont {Tiurev},
  \citenamefont {Ollikainen}, \citenamefont {Kuopanportti}, \citenamefont
  {Nakahara}, \citenamefont {Hall},\ and\ \citenamefont
  {M{\"o}tt{\"o}nen}}]{tiurev2018}%
  \BibitemOpen
  \bibfield  {author} {\bibinfo {author} {\bibfnamefont {K.}~\bibnamefont
  {Tiurev}}, \bibinfo {author} {\bibfnamefont {T.}~\bibnamefont {Ollikainen}},
  \bibinfo {author} {\bibfnamefont {P.}~\bibnamefont {Kuopanportti}}, \bibinfo
  {author} {\bibfnamefont {M.}~\bibnamefont {Nakahara}}, \bibinfo {author}
  {\bibfnamefont {D.~S.}\ \bibnamefont {Hall}}, \ and\ \bibinfo {author}
  {\bibfnamefont {M.}~\bibnamefont {M{\"o}tt{\"o}nen}},\ }\href
  {https://doi.org/10.1088/1367-2630/aac2a8} {\bibfield  {journal} {\bibinfo
  {journal} {New J. Phys.}\ }\textbf {\bibinfo {volume} {20}},\ \bibinfo
  {pages} {055011} (\bibinfo {year} {2018})}\BibitemShut {NoStop}%
\bibitem [{\citenamefont {Leslie}\ \emph
  {et~al.}(2009{\natexlab{a}})\citenamefont {Leslie}, \citenamefont {Hansen},
  \citenamefont {Wright}, \citenamefont {Deutsch},\ and\ \citenamefont
  {Bigelow}}]{leslie2009b}%
  \BibitemOpen
  \bibfield  {author} {\bibinfo {author} {\bibfnamefont {L.~S.}\ \bibnamefont
  {Leslie}}, \bibinfo {author} {\bibfnamefont {A.}~\bibnamefont {Hansen}},
  \bibinfo {author} {\bibfnamefont {K.~C.}\ \bibnamefont {Wright}}, \bibinfo
  {author} {\bibfnamefont {B.~M.}\ \bibnamefont {Deutsch}}, \ and\ \bibinfo
  {author} {\bibfnamefont {N.~P.}\ \bibnamefont {Bigelow}},\ }\href {\doibase
  10.1103/PhysRevLett.103.250401} {\bibfield  {journal} {\bibinfo  {journal}
  {Phys. Rev. Lett.}\ }\textbf {\bibinfo {volume} {103}},\ \bibinfo {pages}
  {250401} (\bibinfo {year} {2009}{\natexlab{a}})}\BibitemShut {NoStop}%
\bibitem [{\citenamefont {Choi}\ \emph {et~al.}(2012)\citenamefont {Choi},
  \citenamefont {Kwon}, \citenamefont {Lee}, \citenamefont {Jeong},
  \citenamefont {An},\ and\ \citenamefont {Shin}}]{choi2012}%
  \BibitemOpen
  \bibfield  {author} {\bibinfo {author} {\bibfnamefont {J.-y.}\ \bibnamefont
  {Choi}}, \bibinfo {author} {\bibfnamefont {W.~J.}\ \bibnamefont {Kwon}},
  \bibinfo {author} {\bibfnamefont {M.}~\bibnamefont {Lee}}, \bibinfo {author}
  {\bibfnamefont {H.}~\bibnamefont {Jeong}}, \bibinfo {author} {\bibfnamefont
  {K.}~\bibnamefont {An}}, \ and\ \bibinfo {author} {\bibfnamefont {Y.-i.}\
  \bibnamefont {Shin}},\ }\href {https://doi.org/10.1088/1367-2630/14/5/053013}
  {\bibfield  {journal} {\bibinfo  {journal} {New J. Phys.}\ }\textbf {\bibinfo
  {volume} {14}},\ \bibinfo {pages} {053013} (\bibinfo {year}
  {2012})}\BibitemShut {NoStop}%
\bibitem [{\citenamefont {Ray}\ \emph {et~al.}(2014)\citenamefont {Ray},
  \citenamefont {Ruokokoski}, \citenamefont {Kandel}, \citenamefont
  {M{\"o}tt{\"o}nen},\ and\ \citenamefont {Hall}}]{ray2014}%
  \BibitemOpen
  \bibfield  {author} {\bibinfo {author} {\bibfnamefont {M.~W.}\ \bibnamefont
  {Ray}}, \bibinfo {author} {\bibfnamefont {E.}~\bibnamefont {Ruokokoski}},
  \bibinfo {author} {\bibfnamefont {S.}~\bibnamefont {Kandel}}, \bibinfo
  {author} {\bibfnamefont {M.}~\bibnamefont {M{\"o}tt{\"o}nen}}, \ and\
  \bibinfo {author} {\bibfnamefont {D.}~\bibnamefont {Hall}},\ }\href {\doibase
  https://doi.org/10.1038/nature12954} {\bibfield  {journal} {\bibinfo
  {journal} {Nature}\ }\textbf {\bibinfo {volume} {505}},\ \bibinfo {pages}
  {657} (\bibinfo {year} {2014})}\BibitemShut {NoStop}%
\bibitem [{\citenamefont {Hall}\ \emph {et~al.}(2016)\citenamefont {Hall},
  \citenamefont {Ray}, \citenamefont {Tiurev}, \citenamefont {Ruokokoski},
  \citenamefont {Gheorghe},\ and\ \citenamefont {M{\"o}tt{\"o}nen}}]{hall2016}%
  \BibitemOpen
  \bibfield  {author} {\bibinfo {author} {\bibfnamefont {D.~S.}\ \bibnamefont
  {Hall}}, \bibinfo {author} {\bibfnamefont {M.~W.}\ \bibnamefont {Ray}},
  \bibinfo {author} {\bibfnamefont {K.}~\bibnamefont {Tiurev}}, \bibinfo
  {author} {\bibfnamefont {E.}~\bibnamefont {Ruokokoski}}, \bibinfo {author}
  {\bibfnamefont {A.~H.}\ \bibnamefont {Gheorghe}}, \ and\ \bibinfo {author}
  {\bibfnamefont {M.}~\bibnamefont {M{\"o}tt{\"o}nen}},\ }\href
  {https://doi.org/10.1038/nphys3624} {\bibfield  {journal} {\bibinfo
  {journal} {Nature Phys.}\ }\textbf {\bibinfo {volume} {12}},\ \bibinfo
  {pages} {478} (\bibinfo {year} {2016})}\BibitemShut {NoStop}%
\bibitem [{\citenamefont {Ollikainen}\ \emph {et~al.}(2017)\citenamefont
  {Ollikainen}, \citenamefont {Tiurev}, \citenamefont {Blinova}, \citenamefont
  {Lee}, \citenamefont {Hall},\ and\ \citenamefont
  {M\"ott\"onen}}]{ollikainen2017}%
  \BibitemOpen
  \bibfield  {author} {\bibinfo {author} {\bibfnamefont {T.}~\bibnamefont
  {Ollikainen}}, \bibinfo {author} {\bibfnamefont {K.}~\bibnamefont {Tiurev}},
  \bibinfo {author} {\bibfnamefont {A.}~\bibnamefont {Blinova}}, \bibinfo
  {author} {\bibfnamefont {W.}~\bibnamefont {Lee}}, \bibinfo {author}
  {\bibfnamefont {D.~S.}\ \bibnamefont {Hall}}, \ and\ \bibinfo {author}
  {\bibfnamefont {M.}~\bibnamefont {M\"ott\"onen}},\ }\href {\doibase
  10.1103/PhysRevX.7.021023} {\bibfield  {journal} {\bibinfo  {journal} {Phys.
  Rev. X}\ }\textbf {\bibinfo {volume} {7}},\ \bibinfo {pages} {021023}
  (\bibinfo {year} {2017})}\BibitemShut {NoStop}%
\bibitem [{\citenamefont {Lee}\ \emph {et~al.}(2018)\citenamefont {Lee},
  \citenamefont {Gheorghe}, \citenamefont {Tiurev}, \citenamefont {Ollikainen},
  \citenamefont {M{\"o}tt{\"o}nen},\ and\ \citenamefont {Hall}}]{lee2018}%
  \BibitemOpen
  \bibfield  {author} {\bibinfo {author} {\bibfnamefont {W.}~\bibnamefont
  {Lee}}, \bibinfo {author} {\bibfnamefont {A.~H.}\ \bibnamefont {Gheorghe}},
  \bibinfo {author} {\bibfnamefont {K.}~\bibnamefont {Tiurev}}, \bibinfo
  {author} {\bibfnamefont {T.}~\bibnamefont {Ollikainen}}, \bibinfo {author}
  {\bibfnamefont {M.}~\bibnamefont {M{\"o}tt{\"o}nen}}, \ and\ \bibinfo
  {author} {\bibfnamefont {D.~S.}\ \bibnamefont {Hall}},\ }\href
  {https://doi.org/10.1126/sciadv.aao3820} {\bibfield  {journal} {\bibinfo
  {journal} {Science advances}\ }\textbf {\bibinfo {volume} {4}},\ \bibinfo
  {pages} {eaao3820} (\bibinfo {year} {2018})}\BibitemShut {NoStop}%
\bibitem [{\citenamefont {Ollikainen}\ \emph {et~al.}(2019)\citenamefont
  {Ollikainen}, \citenamefont {Blinova}, \citenamefont {M\"ott\"onen},\ and\
  \citenamefont {Hall}}]{ollikainen2019}%
  \BibitemOpen
  \bibfield  {author} {\bibinfo {author} {\bibfnamefont {T.}~\bibnamefont
  {Ollikainen}}, \bibinfo {author} {\bibfnamefont {A.}~\bibnamefont {Blinova}},
  \bibinfo {author} {\bibfnamefont {M.}~\bibnamefont {M\"ott\"onen}}, \ and\
  \bibinfo {author} {\bibfnamefont {D.~S.}\ \bibnamefont {Hall}},\ }\href
  {\doibase 10.1103/PhysRevLett.123.163003} {\bibfield  {journal} {\bibinfo
  {journal} {Phys. Rev. Lett.}\ }\textbf {\bibinfo {volume} {123}},\ \bibinfo
  {pages} {163003} (\bibinfo {year} {2019})}\BibitemShut {NoStop}%
\bibitem [{\citenamefont {Weiss}\ \emph {et~al.}(2019)\citenamefont {Weiss},
  \citenamefont {Borgh}, \citenamefont {Blinova}, \citenamefont {Ollikainen},
  \citenamefont {M{\"o}tt{\"o}nen}, \citenamefont {Ruostekoski},\ and\
  \citenamefont {Hall}}]{weiss2019}%
  \BibitemOpen
  \bibfield  {author} {\bibinfo {author} {\bibfnamefont {L.~S.}\ \bibnamefont
  {Weiss}}, \bibinfo {author} {\bibfnamefont {M.~O.}\ \bibnamefont {Borgh}},
  \bibinfo {author} {\bibfnamefont {A.}~\bibnamefont {Blinova}}, \bibinfo
  {author} {\bibfnamefont {T.}~\bibnamefont {Ollikainen}}, \bibinfo {author}
  {\bibfnamefont {M.}~\bibnamefont {M{\"o}tt{\"o}nen}}, \bibinfo {author}
  {\bibfnamefont {J.}~\bibnamefont {Ruostekoski}}, \ and\ \bibinfo {author}
  {\bibfnamefont {D.~S.}\ \bibnamefont {Hall}},\ }\href
  {https://doi.org/10.1038/s41467-019-12787-1} {\bibfield  {journal} {\bibinfo
  {journal} {Nat. Commun.}\ }\textbf {\bibinfo {volume} {10}},\ \bibinfo
  {pages} {4772} (\bibinfo {year} {2019})}\BibitemShut {NoStop}%
\bibitem [{\citenamefont {Underwood}\ \emph {et~al.}(2020)\citenamefont
  {Underwood}, \citenamefont {Baillie}, \citenamefont {Blakie},\ and\
  \citenamefont {Takeuchi}}]{underwood2020}%
  \BibitemOpen
  \bibfield  {author} {\bibinfo {author} {\bibfnamefont {A.~P.~C.}\
  \bibnamefont {Underwood}}, \bibinfo {author} {\bibfnamefont {D.}~\bibnamefont
  {Baillie}}, \bibinfo {author} {\bibfnamefont {P.~B.}\ \bibnamefont {Blakie}},
  \ and\ \bibinfo {author} {\bibfnamefont {H.}~\bibnamefont {Takeuchi}},\
  }\href {\doibase 10.1103/PhysRevA.102.023326} {\bibfield  {journal} {\bibinfo
   {journal} {Phys. Rev. A}\ }\textbf {\bibinfo {volume} {102}},\ \bibinfo
  {pages} {023326} (\bibinfo {year} {2020})}\BibitemShut {NoStop}%
\bibitem [{\citenamefont {Takeuchi}(2020)}]{takeuchi2020}%
  \BibitemOpen
  \bibfield  {author} {\bibinfo {author} {\bibfnamefont {H.}~\bibnamefont
  {Takeuchi}},\ }\href@noop {} {\bibfield  {journal} {\bibinfo  {journal}
  {arXiv:2009.03556}\ } (\bibinfo {year} {2020})}\BibitemShut {NoStop}%
\bibitem [{\citenamefont {Eto}\ \emph {et~al.}(2011)\citenamefont {Eto},
  \citenamefont {Kasamatsu}, \citenamefont {Nitta}, \citenamefont {Takeuchi},\
  and\ \citenamefont {Tsubota}}]{eto2011}%
  \BibitemOpen
  \bibfield  {author} {\bibinfo {author} {\bibfnamefont {M.}~\bibnamefont
  {Eto}}, \bibinfo {author} {\bibfnamefont {K.}~\bibnamefont {Kasamatsu}},
  \bibinfo {author} {\bibfnamefont {M.}~\bibnamefont {Nitta}}, \bibinfo
  {author} {\bibfnamefont {H.}~\bibnamefont {Takeuchi}}, \ and\ \bibinfo
  {author} {\bibfnamefont {M.}~\bibnamefont {Tsubota}},\ }\href {\doibase
  10.1103/PhysRevA.83.063603} {\bibfield  {journal} {\bibinfo  {journal} {Phys.
  Rev. A}\ }\textbf {\bibinfo {volume} {83}},\ \bibinfo {pages} {063603}
  (\bibinfo {year} {2011})}\BibitemShut {NoStop}%
\bibitem [{\citenamefont {Seo}\ \emph {et~al.}(2015)\citenamefont {Seo},
  \citenamefont {Kang}, \citenamefont {Kwon},\ and\ \citenamefont
  {Shin}}]{seo2015}%
  \BibitemOpen
  \bibfield  {author} {\bibinfo {author} {\bibfnamefont {S.~W.}\ \bibnamefont
  {Seo}}, \bibinfo {author} {\bibfnamefont {S.}~\bibnamefont {Kang}}, \bibinfo
  {author} {\bibfnamefont {W.~J.}\ \bibnamefont {Kwon}}, \ and\ \bibinfo
  {author} {\bibfnamefont {Y.-i.}\ \bibnamefont {Shin}},\ }\href {\doibase
  10.1103/PhysRevLett.115.015301} {\bibfield  {journal} {\bibinfo  {journal}
  {Phys. Rev. Lett.}\ }\textbf {\bibinfo {volume} {115}},\ \bibinfo {pages}
  {015301} (\bibinfo {year} {2015})}\BibitemShut {NoStop}%
\bibitem [{\citenamefont {Seo}\ \emph {et~al.}(2016)\citenamefont {Seo},
  \citenamefont {Kwon}, \citenamefont {Kang},\ and\ \citenamefont
  {Shin}}]{seo2016}%
  \BibitemOpen
  \bibfield  {author} {\bibinfo {author} {\bibfnamefont {S.~W.}\ \bibnamefont
  {Seo}}, \bibinfo {author} {\bibfnamefont {W.~J.}\ \bibnamefont {Kwon}},
  \bibinfo {author} {\bibfnamefont {S.}~\bibnamefont {Kang}}, \ and\ \bibinfo
  {author} {\bibfnamefont {Y.}~\bibnamefont {Shin}},\ }\href {\doibase
  10.1103/PhysRevLett.116.185301} {\bibfield  {journal} {\bibinfo  {journal}
  {Phys. Rev. Lett.}\ }\textbf {\bibinfo {volume} {116}},\ \bibinfo {pages}
  {185301} (\bibinfo {year} {2016})}\BibitemShut {NoStop}%
\bibitem [{\citenamefont {Kasamatsu}\ \emph {et~al.}(2016)\citenamefont
  {Kasamatsu}, \citenamefont {Eto},\ and\ \citenamefont
  {Nitta}}]{kasamatsu2016}%
  \BibitemOpen
  \bibfield  {author} {\bibinfo {author} {\bibfnamefont {K.}~\bibnamefont
  {Kasamatsu}}, \bibinfo {author} {\bibfnamefont {M.}~\bibnamefont {Eto}}, \
  and\ \bibinfo {author} {\bibfnamefont {M.}~\bibnamefont {Nitta}},\ }\href
  {\doibase 10.1103/PhysRevA.93.013615} {\bibfield  {journal} {\bibinfo
  {journal} {Phys. Rev. A}\ }\textbf {\bibinfo {volume} {93}},\ \bibinfo
  {pages} {013615} (\bibinfo {year} {2016})}\BibitemShut {NoStop}%
\bibitem [{\citenamefont {Kobayashi}\ \emph {et~al.}(2009)\citenamefont
  {Kobayashi}, \citenamefont {Kawaguchi}, \citenamefont {Nitta},\ and\
  \citenamefont {Ueda}}]{kobayashi2009}%
  \BibitemOpen
  \bibfield  {author} {\bibinfo {author} {\bibfnamefont {M.}~\bibnamefont
  {Kobayashi}}, \bibinfo {author} {\bibfnamefont {Y.}~\bibnamefont
  {Kawaguchi}}, \bibinfo {author} {\bibfnamefont {M.}~\bibnamefont {Nitta}}, \
  and\ \bibinfo {author} {\bibfnamefont {M.}~\bibnamefont {Ueda}},\ }\href
  {\doibase 10.1103/PhysRevLett.103.115301} {\bibfield  {journal} {\bibinfo
  {journal} {Phys. Rev. Lett.}\ }\textbf {\bibinfo {volume} {103}},\ \bibinfo
  {pages} {115301} (\bibinfo {year} {2009})}\BibitemShut {NoStop}%
\bibitem [{\citenamefont {Mawson}\ \emph {et~al.}(2015)\citenamefont {Mawson},
  \citenamefont {Ruben},\ and\ \citenamefont {Simula}}]{mawson2015}%
  \BibitemOpen
  \bibfield  {author} {\bibinfo {author} {\bibfnamefont {T.}~\bibnamefont
  {Mawson}}, \bibinfo {author} {\bibfnamefont {G.}~\bibnamefont {Ruben}}, \
  and\ \bibinfo {author} {\bibfnamefont {T.}~\bibnamefont {Simula}},\ }\href
  {\doibase 10.1103/PhysRevA.91.063630} {\bibfield  {journal} {\bibinfo
  {journal} {Phys. Rev. A}\ }\textbf {\bibinfo {volume} {91}},\ \bibinfo
  {pages} {063630} (\bibinfo {year} {2015})}\BibitemShut {NoStop}%
\bibitem [{\citenamefont {Kaneda}\ and\ \citenamefont
  {Saito}(2014)}]{kaneda2014}%
  \BibitemOpen
  \bibfield  {author} {\bibinfo {author} {\bibfnamefont {T.}~\bibnamefont
  {Kaneda}}\ and\ \bibinfo {author} {\bibfnamefont {H.}~\bibnamefont {Saito}},\
  }\href {\doibase 10.1103/PhysRevA.90.053632} {\bibfield  {journal} {\bibinfo
  {journal} {Phys. Rev. A}\ }\textbf {\bibinfo {volume} {90}},\ \bibinfo
  {pages} {053632} (\bibinfo {year} {2014})}\BibitemShut {NoStop}%
\bibitem [{\citenamefont {Sadler}\ \emph {et~al.}(2006)\citenamefont {Sadler},
  \citenamefont {Higbie}, \citenamefont {Leslie}, \citenamefont
  {Vengalattore},\ and\ \citenamefont {Stamper-Kurn}}]{Sadler2006a}%
  \BibitemOpen
  \bibfield  {author} {\bibinfo {author} {\bibfnamefont {L.~E.}\ \bibnamefont
  {Sadler}}, \bibinfo {author} {\bibfnamefont {J.~M.}\ \bibnamefont {Higbie}},
  \bibinfo {author} {\bibfnamefont {S.~R.}\ \bibnamefont {Leslie}}, \bibinfo
  {author} {\bibfnamefont {M.}~\bibnamefont {Vengalattore}}, \ and\ \bibinfo
  {author} {\bibfnamefont {D.~M.}\ \bibnamefont {Stamper-Kurn}},\ }\href
  {http://dx.doi.org/10.1038/nature05094} {\bibfield  {journal} {\bibinfo
  {journal} {Nature}\ }\textbf {\bibinfo {volume} {443}},\ \bibinfo {pages}
  {312} (\bibinfo {year} {2006})}\BibitemShut {NoStop}%
\bibitem [{\citenamefont {Saito}\ \emph
  {et~al.}(2007{\natexlab{a}})\citenamefont {Saito}, \citenamefont
  {Kawaguchi},\ and\ \citenamefont {Ueda}}]{Saito2007b}%
  \BibitemOpen
  \bibfield  {author} {\bibinfo {author} {\bibfnamefont {H.}~\bibnamefont
  {Saito}}, \bibinfo {author} {\bibfnamefont {Y.}~\bibnamefont {Kawaguchi}}, \
  and\ \bibinfo {author} {\bibfnamefont {M.}~\bibnamefont {Ueda}},\ }\href
  {\doibase 10.1103/PhysRevA.75.013621} {\bibfield  {journal} {\bibinfo
  {journal} {Phys. Rev. A}\ }\textbf {\bibinfo {volume} {75}},\ \bibinfo
  {pages} {013621} (\bibinfo {year} {2007}{\natexlab{a}})}\BibitemShut
  {NoStop}%
\bibitem [{\citenamefont {Saito}\ \emph
  {et~al.}(2007{\natexlab{b}})\citenamefont {Saito}, \citenamefont
  {Kawaguchi},\ and\ \citenamefont {Ueda}}]{Saito2007a}%
  \BibitemOpen
  \bibfield  {author} {\bibinfo {author} {\bibfnamefont {H.}~\bibnamefont
  {Saito}}, \bibinfo {author} {\bibfnamefont {Y.}~\bibnamefont {Kawaguchi}}, \
  and\ \bibinfo {author} {\bibfnamefont {M.}~\bibnamefont {Ueda}},\ }\href
  {\doibase 10.1103/PhysRevA.76.043613} {\bibfield  {journal} {\bibinfo
  {journal} {Phys. Rev. A}\ }\textbf {\bibinfo {volume} {76}},\ \bibinfo
  {pages} {043613} (\bibinfo {year} {2007}{\natexlab{b}})}\BibitemShut
  {NoStop}%
\bibitem [{\citenamefont {Uhlmann}\ \emph {et~al.}(2007)\citenamefont
  {Uhlmann}, \citenamefont {Sch\"utzhold},\ and\ \citenamefont
  {Fischer}}]{uhlmann2007}%
  \BibitemOpen
  \bibfield  {author} {\bibinfo {author} {\bibfnamefont {M.}~\bibnamefont
  {Uhlmann}}, \bibinfo {author} {\bibfnamefont {R.}~\bibnamefont
  {Sch\"utzhold}}, \ and\ \bibinfo {author} {\bibfnamefont {U.~R.}\
  \bibnamefont {Fischer}},\ }\href {\doibase 10.1103/PhysRevLett.99.120407}
  {\bibfield  {journal} {\bibinfo  {journal} {Phys. Rev. Lett.}\ }\textbf
  {\bibinfo {volume} {99}},\ \bibinfo {pages} {120407} (\bibinfo {year}
  {2007})}\BibitemShut {NoStop}%
\bibitem [{\citenamefont {Lamacraft}(2007)}]{lamacraft2007}%
  \BibitemOpen
  \bibfield  {author} {\bibinfo {author} {\bibfnamefont {A.}~\bibnamefont
  {Lamacraft}},\ }\href {\doibase 10.1103/PhysRevLett.98.160404} {\bibfield
  {journal} {\bibinfo  {journal} {Phys. Rev. Lett.}\ }\textbf {\bibinfo
  {volume} {98}},\ \bibinfo {pages} {160404} (\bibinfo {year}
  {2007})}\BibitemShut {NoStop}%
\bibitem [{\citenamefont {Williamson}\ and\ \citenamefont
  {Blakie}(2016{\natexlab{a}})}]{williamson2016a}%
  \BibitemOpen
  \bibfield  {author} {\bibinfo {author} {\bibfnamefont {L.~A.}\ \bibnamefont
  {Williamson}}\ and\ \bibinfo {author} {\bibfnamefont {P.~B.}\ \bibnamefont
  {Blakie}},\ }\href {\doibase 10.1103/PhysRevLett.116.025301} {\bibfield
  {journal} {\bibinfo  {journal} {Phys. Rev. Lett.}\ }\textbf {\bibinfo
  {volume} {116}},\ \bibinfo {pages} {025301} (\bibinfo {year}
  {2016}{\natexlab{a}})}\BibitemShut {NoStop}%
\bibitem [{\citenamefont {Williamson}\ and\ \citenamefont
  {Blakie}(2016{\natexlab{b}})}]{williamson2016b}%
  \BibitemOpen
  \bibfield  {author} {\bibinfo {author} {\bibfnamefont {L.~A.}\ \bibnamefont
  {Williamson}}\ and\ \bibinfo {author} {\bibfnamefont {P.~B.}\ \bibnamefont
  {Blakie}},\ }\href {\doibase 10.1103/PhysRevA.94.023608} {\bibfield
  {journal} {\bibinfo  {journal} {Phys. Rev. A}\ }\textbf {\bibinfo {volume}
  {94}},\ \bibinfo {pages} {023608} (\bibinfo {year}
  {2016}{\natexlab{b}})}\BibitemShut {NoStop}%
\bibitem [{\citenamefont {Schmied}\ \emph {et~al.}(2019)\citenamefont
  {Schmied}, \citenamefont {Gasenzer},\ and\ \citenamefont
  {Blakie}}]{schmied2019}%
  \BibitemOpen
  \bibfield  {author} {\bibinfo {author} {\bibfnamefont {C.-M.}\ \bibnamefont
  {Schmied}}, \bibinfo {author} {\bibfnamefont {T.}~\bibnamefont {Gasenzer}}, \
  and\ \bibinfo {author} {\bibfnamefont {P.~B.}\ \bibnamefont {Blakie}},\
  }\href {\doibase 10.1103/PhysRevA.100.033603} {\bibfield  {journal} {\bibinfo
   {journal} {Phys. Rev. A}\ }\textbf {\bibinfo {volume} {100}},\ \bibinfo
  {pages} {033603} (\bibinfo {year} {2019})}\BibitemShut {NoStop}%
\bibitem [{\citenamefont {Kudo}\ and\ \citenamefont
  {Kawaguchi}(2015)}]{kudo2015}%
  \BibitemOpen
  \bibfield  {author} {\bibinfo {author} {\bibfnamefont {K.}~\bibnamefont
  {Kudo}}\ and\ \bibinfo {author} {\bibfnamefont {Y.}~\bibnamefont
  {Kawaguchi}},\ }\href {\doibase 10.1103/PhysRevA.91.053609} {\bibfield
  {journal} {\bibinfo  {journal} {Phys. Rev. A}\ }\textbf {\bibinfo {volume}
  {91}},\ \bibinfo {pages} {053609} (\bibinfo {year} {2015})}\BibitemShut
  {NoStop}%
\bibitem [{\citenamefont {Aref}(2007)}]{aref2007}%
  \BibitemOpen
  \bibfield  {author} {\bibinfo {author} {\bibfnamefont {H.}~\bibnamefont
  {Aref}},\ }\href {\doibase 10.1063/1.2425103} {\bibfield  {journal} {\bibinfo
   {journal} {J. Math. Phys.}\ }\textbf {\bibinfo {volume} {48}},\ \bibinfo
  {pages} {065401} (\bibinfo {year} {2007})}\BibitemShut {NoStop}%
\bibitem [{\citenamefont {Gauthier}\ \emph {et~al.}(2019)\citenamefont
  {Gauthier}, \citenamefont {Reeves}, \citenamefont {Yu}, \citenamefont
  {Bradley}, \citenamefont {Baker}, \citenamefont {Bell}, \citenamefont
  {Rubinsztein-Dunlop}, \citenamefont {Davis},\ and\ \citenamefont
  {Neely}}]{gauthier2019}%
  \BibitemOpen
  \bibfield  {author} {\bibinfo {author} {\bibfnamefont {G.}~\bibnamefont
  {Gauthier}}, \bibinfo {author} {\bibfnamefont {M.~T.}\ \bibnamefont
  {Reeves}}, \bibinfo {author} {\bibfnamefont {X.}~\bibnamefont {Yu}}, \bibinfo
  {author} {\bibfnamefont {A.~S.}\ \bibnamefont {Bradley}}, \bibinfo {author}
  {\bibfnamefont {M.~A.}\ \bibnamefont {Baker}}, \bibinfo {author}
  {\bibfnamefont {T.~A.}\ \bibnamefont {Bell}}, \bibinfo {author}
  {\bibfnamefont {H.}~\bibnamefont {Rubinsztein-Dunlop}}, \bibinfo {author}
  {\bibfnamefont {M.~J.}\ \bibnamefont {Davis}}, \ and\ \bibinfo {author}
  {\bibfnamefont {T.~W.}\ \bibnamefont {Neely}},\ }\href {\doibase
  10.1126/science.aat5718} {\bibfield  {journal} {\bibinfo  {journal}
  {Science}\ }\textbf {\bibinfo {volume} {364}},\ \bibinfo {pages} {1264}
  (\bibinfo {year} {2019})}\BibitemShut {NoStop}%
\bibitem [{\citenamefont {Johnstone}\ \emph {et~al.}(2019)\citenamefont
  {Johnstone}, \citenamefont {Groszek}, \citenamefont {Starkey}, \citenamefont
  {Billington}, \citenamefont {Simula},\ and\ \citenamefont
  {Helmerson}}]{johnstone2019}%
  \BibitemOpen
  \bibfield  {author} {\bibinfo {author} {\bibfnamefont {S.~P.}\ \bibnamefont
  {Johnstone}}, \bibinfo {author} {\bibfnamefont {A.~J.}\ \bibnamefont
  {Groszek}}, \bibinfo {author} {\bibfnamefont {P.~T.}\ \bibnamefont
  {Starkey}}, \bibinfo {author} {\bibfnamefont {C.~J.}\ \bibnamefont
  {Billington}}, \bibinfo {author} {\bibfnamefont {T.~P.}\ \bibnamefont
  {Simula}}, \ and\ \bibinfo {author} {\bibfnamefont {K.}~\bibnamefont
  {Helmerson}},\ }\href@noop {} {\bibfield  {journal} {\bibinfo  {journal}
  {Science}\ }\textbf {\bibinfo {volume} {364}},\ \bibinfo {pages} {1267}
  (\bibinfo {year} {2019})}\BibitemShut {NoStop}%
\bibitem [{\citenamefont {Sachkou}\ \emph {et~al.}(2019)\citenamefont
  {Sachkou}, \citenamefont {Baker}, \citenamefont {Harris}, \citenamefont
  {Stockdale}, \citenamefont {Forstner}, \citenamefont {Reeves}, \citenamefont
  {He}, \citenamefont {McAuslan}, \citenamefont {Bradley}, \citenamefont
  {Davis},\ and\ \citenamefont {Bowen}}]{sachkou2019}%
  \BibitemOpen
  \bibfield  {author} {\bibinfo {author} {\bibfnamefont {Y.~P.}\ \bibnamefont
  {Sachkou}}, \bibinfo {author} {\bibfnamefont {C.~G.}\ \bibnamefont {Baker}},
  \bibinfo {author} {\bibfnamefont {G.~I.}\ \bibnamefont {Harris}}, \bibinfo
  {author} {\bibfnamefont {O.~R.}\ \bibnamefont {Stockdale}}, \bibinfo {author}
  {\bibfnamefont {S.}~\bibnamefont {Forstner}}, \bibinfo {author}
  {\bibfnamefont {M.~T.}\ \bibnamefont {Reeves}}, \bibinfo {author}
  {\bibfnamefont {X.}~\bibnamefont {He}}, \bibinfo {author} {\bibfnamefont
  {D.~L.}\ \bibnamefont {McAuslan}}, \bibinfo {author} {\bibfnamefont {A.~S.}\
  \bibnamefont {Bradley}}, \bibinfo {author} {\bibfnamefont {M.~J.}\
  \bibnamefont {Davis}}, \ and\ \bibinfo {author} {\bibfnamefont {W.~P.}\
  \bibnamefont {Bowen}},\ }\href {\doibase 10.1126/science.aaw9229} {\bibfield
  {journal} {\bibinfo  {journal} {Science}\ }\textbf {\bibinfo {volume}
  {366}},\ \bibinfo {pages} {1480} (\bibinfo {year} {2019})}\BibitemShut
  {NoStop}%
\bibitem [{\citenamefont {Middelkamp}\ \emph {et~al.}(2011)\citenamefont
  {Middelkamp}, \citenamefont {Torres}, \citenamefont {Kevrekidis},
  \citenamefont {Frantzeskakis}, \citenamefont {Carretero-Gonz\'alez},
  \citenamefont {Schmelcher}, \citenamefont {Freilich},\ and\ \citenamefont
  {Hall}}]{middelkamp2011}%
  \BibitemOpen
  \bibfield  {author} {\bibinfo {author} {\bibfnamefont {S.}~\bibnamefont
  {Middelkamp}}, \bibinfo {author} {\bibfnamefont {P.~J.}\ \bibnamefont
  {Torres}}, \bibinfo {author} {\bibfnamefont {P.~G.}\ \bibnamefont
  {Kevrekidis}}, \bibinfo {author} {\bibfnamefont {D.~J.}\ \bibnamefont
  {Frantzeskakis}}, \bibinfo {author} {\bibfnamefont {R.}~\bibnamefont
  {Carretero-Gonz\'alez}}, \bibinfo {author} {\bibfnamefont {P.}~\bibnamefont
  {Schmelcher}}, \bibinfo {author} {\bibfnamefont {D.~V.}\ \bibnamefont
  {Freilich}}, \ and\ \bibinfo {author} {\bibfnamefont {D.~S.}\ \bibnamefont
  {Hall}},\ }\href {\doibase 10.1103/PhysRevA.84.011605} {\bibfield  {journal}
  {\bibinfo  {journal} {Phys. Rev. A}\ }\textbf {\bibinfo {volume} {84}},\
  \bibinfo {pages} {011605} (\bibinfo {year} {2011})}\BibitemShut {NoStop}%
\bibitem [{\citenamefont {Navarro}\ \emph {et~al.}(2013)\citenamefont
  {Navarro}, \citenamefont {Carretero-Gonz\'alez}, \citenamefont {Torres},
  \citenamefont {Kevrekidis}, \citenamefont {Frantzeskakis}, \citenamefont
  {Ray}, \citenamefont {Altunta\ifmmode~\mbox{\c{s}}\else \c{s}\fi{}},\ and\
  \citenamefont {Hall}}]{navarro2013}%
  \BibitemOpen
  \bibfield  {author} {\bibinfo {author} {\bibfnamefont {R.}~\bibnamefont
  {Navarro}}, \bibinfo {author} {\bibfnamefont {R.}~\bibnamefont
  {Carretero-Gonz\'alez}}, \bibinfo {author} {\bibfnamefont {P.~J.}\
  \bibnamefont {Torres}}, \bibinfo {author} {\bibfnamefont {P.~G.}\
  \bibnamefont {Kevrekidis}}, \bibinfo {author} {\bibfnamefont {D.~J.}\
  \bibnamefont {Frantzeskakis}}, \bibinfo {author} {\bibfnamefont {M.~W.}\
  \bibnamefont {Ray}}, \bibinfo {author} {\bibfnamefont {E.}~\bibnamefont
  {Altunta\ifmmode~\mbox{\c{s}}\else \c{s}\fi{}}}, \ and\ \bibinfo {author}
  {\bibfnamefont {D.~S.}\ \bibnamefont {Hall}},\ }\href {\doibase
  10.1103/PhysRevLett.110.225301} {\bibfield  {journal} {\bibinfo  {journal}
  {Phys. Rev. Lett.}\ }\textbf {\bibinfo {volume} {110}},\ \bibinfo {pages}
  {225301} (\bibinfo {year} {2013})}\BibitemShut {NoStop}%
\bibitem [{\citenamefont {Moon}\ \emph {et~al.}(2015)\citenamefont {Moon},
  \citenamefont {Kwon}, \citenamefont {Lee},\ and\ \citenamefont
  {Shin}}]{moon2015}%
  \BibitemOpen
  \bibfield  {author} {\bibinfo {author} {\bibfnamefont {G.}~\bibnamefont
  {Moon}}, \bibinfo {author} {\bibfnamefont {W.~J.}\ \bibnamefont {Kwon}},
  \bibinfo {author} {\bibfnamefont {H.}~\bibnamefont {Lee}}, \ and\ \bibinfo
  {author} {\bibfnamefont {Y.-i.}\ \bibnamefont {Shin}},\ }\href {\doibase
  10.1103/PhysRevA.92.051601} {\bibfield  {journal} {\bibinfo  {journal} {Phys.
  Rev. A}\ }\textbf {\bibinfo {volume} {92}},\ \bibinfo {pages} {051601}
  (\bibinfo {year} {2015})}\BibitemShut {NoStop}%
\bibitem [{\citenamefont {Reeves}\ \emph {et~al.}(2013)\citenamefont {Reeves},
  \citenamefont {Billam}, \citenamefont {Anderson},\ and\ \citenamefont
  {Bradley}}]{reeves2013}%
  \BibitemOpen
  \bibfield  {author} {\bibinfo {author} {\bibfnamefont {M.~T.}\ \bibnamefont
  {Reeves}}, \bibinfo {author} {\bibfnamefont {T.~P.}\ \bibnamefont {Billam}},
  \bibinfo {author} {\bibfnamefont {B.~P.}\ \bibnamefont {Anderson}}, \ and\
  \bibinfo {author} {\bibfnamefont {A.~S.}\ \bibnamefont {Bradley}},\ }\href
  {\doibase 10.1103/PhysRevLett.110.104501} {\bibfield  {journal} {\bibinfo
  {journal} {Phys. Rev. Lett.}\ }\textbf {\bibinfo {volume} {110}},\ \bibinfo
  {pages} {104501} (\bibinfo {year} {2013})}\BibitemShut {NoStop}%
\bibitem [{\citenamefont {Billam}\ \emph {et~al.}(2014)\citenamefont {Billam},
  \citenamefont {Reeves}, \citenamefont {Anderson},\ and\ \citenamefont
  {Bradley}}]{billam2014}%
  \BibitemOpen
  \bibfield  {author} {\bibinfo {author} {\bibfnamefont {T.~P.}\ \bibnamefont
  {Billam}}, \bibinfo {author} {\bibfnamefont {M.~T.}\ \bibnamefont {Reeves}},
  \bibinfo {author} {\bibfnamefont {B.~P.}\ \bibnamefont {Anderson}}, \ and\
  \bibinfo {author} {\bibfnamefont {A.~S.}\ \bibnamefont {Bradley}},\ }\href
  {\doibase 10.1103/PhysRevLett.112.145301} {\bibfield  {journal} {\bibinfo
  {journal} {Phys. Rev. Lett.}\ }\textbf {\bibinfo {volume} {112}},\ \bibinfo
  {pages} {145301} (\bibinfo {year} {2014})}\BibitemShut {NoStop}%
\bibitem [{\citenamefont {Reeves}\ \emph {et~al.}(2014)\citenamefont {Reeves},
  \citenamefont {Billam}, \citenamefont {Anderson},\ and\ \citenamefont
  {Bradley}}]{reeves2014}%
  \BibitemOpen
  \bibfield  {author} {\bibinfo {author} {\bibfnamefont {M.~T.}\ \bibnamefont
  {Reeves}}, \bibinfo {author} {\bibfnamefont {T.~P.}\ \bibnamefont {Billam}},
  \bibinfo {author} {\bibfnamefont {B.~P.}\ \bibnamefont {Anderson}}, \ and\
  \bibinfo {author} {\bibfnamefont {A.~S.}\ \bibnamefont {Bradley}},\ }\href
  {\doibase 10.1103/PhysRevA.89.053631} {\bibfield  {journal} {\bibinfo
  {journal} {Phys. Rev. A}\ }\textbf {\bibinfo {volume} {89}},\ \bibinfo
  {pages} {053631} (\bibinfo {year} {2014})}\BibitemShut {NoStop}%
\bibitem [{\citenamefont {Simula}\ \emph {et~al.}(2014)\citenamefont {Simula},
  \citenamefont {Davis},\ and\ \citenamefont {Helmerson}}]{simula2014}%
  \BibitemOpen
  \bibfield  {author} {\bibinfo {author} {\bibfnamefont {T.}~\bibnamefont
  {Simula}}, \bibinfo {author} {\bibfnamefont {M.~J.}\ \bibnamefont {Davis}}, \
  and\ \bibinfo {author} {\bibfnamefont {K.}~\bibnamefont {Helmerson}},\ }\href
  {\doibase 10.1103/PhysRevLett.113.165302} {\bibfield  {journal} {\bibinfo
  {journal} {Phys. Rev. Lett.}\ }\textbf {\bibinfo {volume} {113}},\ \bibinfo
  {pages} {165302} (\bibinfo {year} {2014})}\BibitemShut {NoStop}%
\bibitem [{\citenamefont {Billam}\ \emph {et~al.}(2015)\citenamefont {Billam},
  \citenamefont {Reeves},\ and\ \citenamefont {Bradley}}]{billam2015}%
  \BibitemOpen
  \bibfield  {author} {\bibinfo {author} {\bibfnamefont {T.~P.}\ \bibnamefont
  {Billam}}, \bibinfo {author} {\bibfnamefont {M.~T.}\ \bibnamefont {Reeves}},
  \ and\ \bibinfo {author} {\bibfnamefont {A.~S.}\ \bibnamefont {Bradley}},\
  }\href {\doibase 10.1103/PhysRevA.91.023615} {\bibfield  {journal} {\bibinfo
  {journal} {Phys. Rev. A}\ }\textbf {\bibinfo {volume} {91}},\ \bibinfo
  {pages} {023615} (\bibinfo {year} {2015})}\BibitemShut {NoStop}%
\bibitem [{\citenamefont {Groszek}\ \emph {et~al.}(2018)\citenamefont
  {Groszek}, \citenamefont {Davis}, \citenamefont {Paganin}, \citenamefont
  {Helmerson},\ and\ \citenamefont {Simula}}]{groszek2018}%
  \BibitemOpen
  \bibfield  {author} {\bibinfo {author} {\bibfnamefont {A.~J.}\ \bibnamefont
  {Groszek}}, \bibinfo {author} {\bibfnamefont {M.~J.}\ \bibnamefont {Davis}},
  \bibinfo {author} {\bibfnamefont {D.~M.}\ \bibnamefont {Paganin}}, \bibinfo
  {author} {\bibfnamefont {K.}~\bibnamefont {Helmerson}}, \ and\ \bibinfo
  {author} {\bibfnamefont {T.~P.}\ \bibnamefont {Simula}},\ }\href {\doibase
  10.1103/PhysRevLett.120.034504} {\bibfield  {journal} {\bibinfo  {journal}
  {Phys. Rev. Lett.}\ }\textbf {\bibinfo {volume} {120}},\ \bibinfo {pages}
  {034504} (\bibinfo {year} {2018})}\BibitemShut {NoStop}%
\bibitem [{\citenamefont {Stockdale}\ \emph {et~al.}(2020)\citenamefont
  {Stockdale}, \citenamefont {Reeves}, \citenamefont {Yu}, \citenamefont
  {Gauthier}, \citenamefont {Goddard-Lee}, \citenamefont {Bowen}, \citenamefont
  {Neely},\ and\ \citenamefont {Davis}}]{stockdale2020}%
  \BibitemOpen
  \bibfield  {author} {\bibinfo {author} {\bibfnamefont {O.~R.}\ \bibnamefont
  {Stockdale}}, \bibinfo {author} {\bibfnamefont {M.~T.}\ \bibnamefont
  {Reeves}}, \bibinfo {author} {\bibfnamefont {X.}~\bibnamefont {Yu}}, \bibinfo
  {author} {\bibfnamefont {G.}~\bibnamefont {Gauthier}}, \bibinfo {author}
  {\bibfnamefont {K.}~\bibnamefont {Goddard-Lee}}, \bibinfo {author}
  {\bibfnamefont {W.~P.}\ \bibnamefont {Bowen}}, \bibinfo {author}
  {\bibfnamefont {T.~W.}\ \bibnamefont {Neely}}, \ and\ \bibinfo {author}
  {\bibfnamefont {M.~J.}\ \bibnamefont {Davis}},\ }\href {\doibase
  10.1103/PhysRevResearch.2.033138} {\bibfield  {journal} {\bibinfo  {journal}
  {Phys. Rev. Research}\ }\textbf {\bibinfo {volume} {2}},\ \bibinfo {pages}
  {033138} (\bibinfo {year} {2020})}\BibitemShut {NoStop}%
\bibitem [{\citenamefont {Turner}(2009)}]{Turner2009}%
  \BibitemOpen
  \bibfield  {author} {\bibinfo {author} {\bibfnamefont {A.~M.}\ \bibnamefont
  {Turner}},\ }\href {\doibase 10.1103/PhysRevLett.103.080603} {\bibfield
  {journal} {\bibinfo  {journal} {Phys. Rev. Lett.}\ }\textbf {\bibinfo
  {volume} {103}},\ \bibinfo {pages} {080603} (\bibinfo {year}
  {2009})}\BibitemShut {NoStop}%
\bibitem [{\citenamefont {Fetter}(1965)}]{fetter1965}%
  \BibitemOpen
  \bibfield  {author} {\bibinfo {author} {\bibfnamefont {A.~L.}\ \bibnamefont
  {Fetter}},\ }\href {\doibase 10.1103/PhysRev.138.A429} {\bibfield  {journal}
  {\bibinfo  {journal} {Phys. Rev.}\ }\textbf {\bibinfo {volume} {138}},\
  \bibinfo {pages} {A429} (\bibinfo {year} {1965})}\BibitemShut {NoStop}%
\bibitem [{\citenamefont {Fetter}(1966)}]{fetter1966}%
  \BibitemOpen
  \bibfield  {author} {\bibinfo {author} {\bibfnamefont {A.~L.}\ \bibnamefont
  {Fetter}},\ }\href {\doibase 10.1103/PhysRev.151.100} {\bibfield  {journal}
  {\bibinfo  {journal} {Phys. Rev.}\ }\textbf {\bibinfo {volume} {151}},\
  \bibinfo {pages} {100} (\bibinfo {year} {1966})}\BibitemShut {NoStop}%
\bibitem [{\citenamefont {Jones}\ and\ \citenamefont
  {Roberts}(1982)}]{jones1982}%
  \BibitemOpen
  \bibfield  {author} {\bibinfo {author} {\bibfnamefont {C.}~\bibnamefont
  {Jones}}\ and\ \bibinfo {author} {\bibfnamefont {P.}~\bibnamefont
  {Roberts}},\ }\href {\doibase 10.1088/0305-4470/15/8/036} {\bibfield
  {journal} {\bibinfo  {journal} {J. Phys. A: Math. Gen.}\ }\textbf {\bibinfo
  {volume} {15}},\ \bibinfo {pages} {2599} (\bibinfo {year}
  {1982})}\BibitemShut {NoStop}%
\bibitem [{\citenamefont {Williamson}\ and\ \citenamefont
  {Blakie}(2016{\natexlab{c}})}]{williamson2016c}%
  \BibitemOpen
  \bibfield  {author} {\bibinfo {author} {\bibfnamefont {L.~A.}\ \bibnamefont
  {Williamson}}\ and\ \bibinfo {author} {\bibfnamefont {P.~B.}\ \bibnamefont
  {Blakie}},\ }\href {\doibase 10.1103/PhysRevA.94.063615} {\bibfield
  {journal} {\bibinfo  {journal} {Phys. Rev. A}\ }\textbf {\bibinfo {volume}
  {94}},\ \bibinfo {pages} {063615} (\bibinfo {year}
  {2016}{\natexlab{c}})}\BibitemShut {NoStop}%
\bibitem [{\citenamefont {Williamson}\ and\ \citenamefont
  {Blakie}(2019)}]{williamson2019}%
  \BibitemOpen
  \bibfield  {author} {\bibinfo {author} {\bibfnamefont {L.~A.}\ \bibnamefont
  {Williamson}}\ and\ \bibinfo {author} {\bibfnamefont {P.~B.}\ \bibnamefont
  {Blakie}},\ }\href {\doibase 10.21468/SciPostPhys.7.3.029} {\bibfield
  {journal} {\bibinfo  {journal} {SciPost Phys.}\ }\textbf {\bibinfo {volume}
  {7}},\ \bibinfo {pages} {29} (\bibinfo {year} {2019})}\BibitemShut {NoStop}%
\bibitem [{\citenamefont {Murata}\ \emph {et~al.}(2007)\citenamefont {Murata},
  \citenamefont {Saito},\ and\ \citenamefont {Ueda}}]{murata2007}%
  \BibitemOpen
  \bibfield  {author} {\bibinfo {author} {\bibfnamefont {K.}~\bibnamefont
  {Murata}}, \bibinfo {author} {\bibfnamefont {H.}~\bibnamefont {Saito}}, \
  and\ \bibinfo {author} {\bibfnamefont {M.}~\bibnamefont {Ueda}},\ }\href
  {\doibase 10.1103/PhysRevA.75.013607} {\bibfield  {journal} {\bibinfo
  {journal} {Phys. Rev. A}\ }\textbf {\bibinfo {volume} {75}},\ \bibinfo
  {pages} {013607} (\bibinfo {year} {2007})}\BibitemShut {NoStop}%
\bibitem [{\citenamefont {Ho}(1998)}]{Ho1998a}%
  \BibitemOpen
  \bibfield  {author} {\bibinfo {author} {\bibfnamefont {T.-L.}\ \bibnamefont
  {Ho}},\ }\href {\doibase 10.1103/PhysRevLett.81.742} {\bibfield  {journal}
  {\bibinfo  {journal} {Phys. Rev. Lett.}\ }\textbf {\bibinfo {volume} {81}},\
  \bibinfo {pages} {742} (\bibinfo {year} {1998})}\BibitemShut {NoStop}%
\bibitem [{\citenamefont {Ohmi}\ and\ \citenamefont
  {Machida}(1998)}]{Ohmi1998a}%
  \BibitemOpen
  \bibfield  {author} {\bibinfo {author} {\bibfnamefont {T.}~\bibnamefont
  {Ohmi}}\ and\ \bibinfo {author} {\bibfnamefont {K.}~\bibnamefont {Machida}},\
  }\href {\doibase 10.1143/JPSJ.67.1822} {\bibfield  {journal} {\bibinfo
  {journal} {J. Phys. Soc. Jpn}\ }\textbf {\bibinfo {volume} {67}},\ \bibinfo
  {pages} {1822} (\bibinfo {year} {1998})}\BibitemShut {NoStop}%
\bibitem [{\citenamefont {Barnett}\ \emph {et~al.}(2011)\citenamefont
  {Barnett}, \citenamefont {Polkovnikov},\ and\ \citenamefont
  {Vengalattore}}]{barnett2011}%
  \BibitemOpen
  \bibfield  {author} {\bibinfo {author} {\bibfnamefont {R.}~\bibnamefont
  {Barnett}}, \bibinfo {author} {\bibfnamefont {A.}~\bibnamefont
  {Polkovnikov}}, \ and\ \bibinfo {author} {\bibfnamefont {M.}~\bibnamefont
  {Vengalattore}},\ }\href {\doibase 10.1103/PhysRevA.84.023606} {\bibfield
  {journal} {\bibinfo  {journal} {Phys. Rev. A}\ }\textbf {\bibinfo {volume}
  {84}},\ \bibinfo {pages} {023606} (\bibinfo {year} {2011})}\BibitemShut
  {NoStop}%
\bibitem [{\citenamefont {Schmaljohann}\ \emph {et~al.}(2004)\citenamefont
  {Schmaljohann}, \citenamefont {Erhard}, \citenamefont {Kronj\"ager},
  \citenamefont {Kottke}, \citenamefont {van Staa}, \citenamefont
  {Cacciapuoti}, \citenamefont {Arlt}, \citenamefont {Bongs},\ and\
  \citenamefont {Sengstock}}]{schmaljohann2004}%
  \BibitemOpen
  \bibfield  {author} {\bibinfo {author} {\bibfnamefont {H.}~\bibnamefont
  {Schmaljohann}}, \bibinfo {author} {\bibfnamefont {M.}~\bibnamefont
  {Erhard}}, \bibinfo {author} {\bibfnamefont {J.}~\bibnamefont {Kronj\"ager}},
  \bibinfo {author} {\bibfnamefont {M.}~\bibnamefont {Kottke}}, \bibinfo
  {author} {\bibfnamefont {S.}~\bibnamefont {van Staa}}, \bibinfo {author}
  {\bibfnamefont {L.}~\bibnamefont {Cacciapuoti}}, \bibinfo {author}
  {\bibfnamefont {J.~J.}\ \bibnamefont {Arlt}}, \bibinfo {author}
  {\bibfnamefont {K.}~\bibnamefont {Bongs}}, \ and\ \bibinfo {author}
  {\bibfnamefont {K.}~\bibnamefont {Sengstock}},\ }\href {\doibase
  10.1103/PhysRevLett.92.040402} {\bibfield  {journal} {\bibinfo  {journal}
  {Phys. Rev. Lett.}\ }\textbf {\bibinfo {volume} {92}},\ \bibinfo {pages}
  {040402} (\bibinfo {year} {2004})}\BibitemShut {NoStop}%
\bibitem [{\citenamefont {Chang}\ \emph {et~al.}(2004)\citenamefont {Chang},
  \citenamefont {Hamley}, \citenamefont {Barrett}, \citenamefont {Sauer},
  \citenamefont {Fortier}, \citenamefont {Zhang}, \citenamefont {You},\ and\
  \citenamefont {Chapman}}]{Chang2004a}%
  \BibitemOpen
  \bibfield  {author} {\bibinfo {author} {\bibfnamefont {M.-S.}\ \bibnamefont
  {Chang}}, \bibinfo {author} {\bibfnamefont {C.~D.}\ \bibnamefont {Hamley}},
  \bibinfo {author} {\bibfnamefont {M.~D.}\ \bibnamefont {Barrett}}, \bibinfo
  {author} {\bibfnamefont {J.~A.}\ \bibnamefont {Sauer}}, \bibinfo {author}
  {\bibfnamefont {K.~M.}\ \bibnamefont {Fortier}}, \bibinfo {author}
  {\bibfnamefont {W.}~\bibnamefont {Zhang}}, \bibinfo {author} {\bibfnamefont
  {L.}~\bibnamefont {You}}, \ and\ \bibinfo {author} {\bibfnamefont {M.~S.}\
  \bibnamefont {Chapman}},\ }\href {\doibase 10.1103/PhysRevLett.92.140403}
  {\bibfield  {journal} {\bibinfo  {journal} {Phys. Rev. Lett.}\ }\textbf
  {\bibinfo {volume} {92}},\ \bibinfo {pages} {140403} (\bibinfo {year}
  {2004})}\BibitemShut {NoStop}%
\bibitem [{\citenamefont {Huh}\ \emph {et~al.}(2020)\citenamefont {Huh},
  \citenamefont {Kim}, \citenamefont {Kwon},\ and\ \citenamefont
  {Choi}}]{huh2020}%
  \BibitemOpen
  \bibfield  {author} {\bibinfo {author} {\bibfnamefont {S.}~\bibnamefont
  {Huh}}, \bibinfo {author} {\bibfnamefont {K.}~\bibnamefont {Kim}}, \bibinfo
  {author} {\bibfnamefont {K.}~\bibnamefont {Kwon}}, \ and\ \bibinfo {author}
  {\bibfnamefont {J.-y.}\ \bibnamefont {Choi}},\ }\href {\doibase
  10.1103/PhysRevResearch.2.033471} {\bibfield  {journal} {\bibinfo  {journal}
  {Phys. Rev. Research}\ }\textbf {\bibinfo {volume} {2}},\ \bibinfo {pages}
  {033471} (\bibinfo {year} {2020})}\BibitemShut {NoStop}%
\bibitem [{\citenamefont {Gerbier}\ \emph {et~al.}(2006)\citenamefont
  {Gerbier}, \citenamefont {Widera}, \citenamefont {F\"olling}, \citenamefont
  {Mandel},\ and\ \citenamefont {Bloch}}]{gerbier2006}%
  \BibitemOpen
  \bibfield  {author} {\bibinfo {author} {\bibfnamefont {F.}~\bibnamefont
  {Gerbier}}, \bibinfo {author} {\bibfnamefont {A.}~\bibnamefont {Widera}},
  \bibinfo {author} {\bibfnamefont {S.}~\bibnamefont {F\"olling}}, \bibinfo
  {author} {\bibfnamefont {O.}~\bibnamefont {Mandel}}, \ and\ \bibinfo {author}
  {\bibfnamefont {I.}~\bibnamefont {Bloch}},\ }\href {\doibase
  10.1103/PhysRevA.73.041602} {\bibfield  {journal} {\bibinfo  {journal} {Phys.
  Rev. A}\ }\textbf {\bibinfo {volume} {73}},\ \bibinfo {pages} {041602}
  (\bibinfo {year} {2006})}\BibitemShut {NoStop}%
\bibitem [{\citenamefont {Leslie}\ \emph
  {et~al.}(2009{\natexlab{b}})\citenamefont {Leslie}, \citenamefont {Guzman},
  \citenamefont {Vengalattore}, \citenamefont {Sau}, \citenamefont {Cohen},\
  and\ \citenamefont {Stamper-Kurn}}]{leslie2009}%
  \BibitemOpen
  \bibfield  {author} {\bibinfo {author} {\bibfnamefont {S.~R.}\ \bibnamefont
  {Leslie}}, \bibinfo {author} {\bibfnamefont {J.}~\bibnamefont {Guzman}},
  \bibinfo {author} {\bibfnamefont {M.}~\bibnamefont {Vengalattore}}, \bibinfo
  {author} {\bibfnamefont {J.~D.}\ \bibnamefont {Sau}}, \bibinfo {author}
  {\bibfnamefont {M.~L.}\ \bibnamefont {Cohen}}, \ and\ \bibinfo {author}
  {\bibfnamefont {D.~M.}\ \bibnamefont {Stamper-Kurn}},\ }\href {\doibase
  10.1103/PhysRevA.79.043631} {\bibfield  {journal} {\bibinfo  {journal} {Phys.
  Rev. A}\ }\textbf {\bibinfo {volume} {79}},\ \bibinfo {pages} {043631}
  (\bibinfo {year} {2009}{\natexlab{b}})}\BibitemShut {NoStop}%
\bibitem [{\citenamefont {Stenger}\ \emph {et~al.}(1998)\citenamefont
  {Stenger}, \citenamefont {Inouye}, \citenamefont {Stamper-Kurn},
  \citenamefont {Miesner}, \citenamefont {Chikkatur},\ and\ \citenamefont
  {Ketterle}}]{Stenger1998a}%
  \BibitemOpen
  \bibfield  {author} {\bibinfo {author} {\bibfnamefont {J.}~\bibnamefont
  {Stenger}}, \bibinfo {author} {\bibfnamefont {S.}~\bibnamefont {Inouye}},
  \bibinfo {author} {\bibfnamefont {D.~M.}\ \bibnamefont {Stamper-Kurn}},
  \bibinfo {author} {\bibfnamefont {H.-J.}\ \bibnamefont {Miesner}}, \bibinfo
  {author} {\bibfnamefont {A.~P.}\ \bibnamefont {Chikkatur}}, \ and\ \bibinfo
  {author} {\bibfnamefont {W.}~\bibnamefont {Ketterle}},\ }\href {\doibase
  10.1038/24567} {\bibfield  {journal} {\bibinfo  {journal} {Nature}\ }\textbf
  {\bibinfo {volume} {396}},\ \bibinfo {pages} {345} (\bibinfo {year}
  {1998})}\BibitemShut {NoStop}%
\bibitem [{\citenamefont {Yukawa}\ and\ \citenamefont
  {Ueda}(2012)}]{Yukawa2012}%
  \BibitemOpen
  \bibfield  {author} {\bibinfo {author} {\bibfnamefont {E.}~\bibnamefont
  {Yukawa}}\ and\ \bibinfo {author} {\bibfnamefont {M.}~\bibnamefont {Ueda}},\
  }\href {\doibase 10.1103/PhysRevA.86.063614} {\bibfield  {journal} {\bibinfo
  {journal} {Phys. Rev. A}\ }\textbf {\bibinfo {volume} {86}},\ \bibinfo
  {pages} {063614} (\bibinfo {year} {2012})}\BibitemShut {NoStop}%
\bibitem [{\citenamefont {Lucas}\ and\ \citenamefont
  {Sur\'owka}(2014)}]{lucas2014}%
  \BibitemOpen
  \bibfield  {author} {\bibinfo {author} {\bibfnamefont {A.}~\bibnamefont
  {Lucas}}\ and\ \bibinfo {author} {\bibfnamefont {P.}~\bibnamefont
  {Sur\'owka}},\ }\href {\doibase 10.1103/PhysRevA.90.053617} {\bibfield
  {journal} {\bibinfo  {journal} {Phys. Rev. A}\ }\textbf {\bibinfo {volume}
  {90}},\ \bibinfo {pages} {053617} (\bibinfo {year} {2014})}\BibitemShut
  {NoStop}%
\bibitem [{\citenamefont {Lin}\ and\ \citenamefont {Xin}(1999)}]{lin1999}%
  \BibitemOpen
  \bibfield  {author} {\bibinfo {author} {\bibfnamefont {F.}~\bibnamefont
  {Lin}}\ and\ \bibinfo {author} {\bibfnamefont {J.}~\bibnamefont {Xin}},\
  }\href {https://doi.org/10.1007/s002200050529} {\bibfield  {journal}
  {\bibinfo  {journal} {Comm. Math. Phys.}\ }\textbf {\bibinfo {volume}
  {200}},\ \bibinfo {pages} {249} (\bibinfo {year} {1999})}\BibitemShut
  {NoStop}%
\bibitem [{\citenamefont {Thouless}\ \emph {et~al.}(1996)\citenamefont
  {Thouless}, \citenamefont {Ao},\ and\ \citenamefont {Niu}}]{thouless1996}%
  \BibitemOpen
  \bibfield  {author} {\bibinfo {author} {\bibfnamefont {D.~J.}\ \bibnamefont
  {Thouless}}, \bibinfo {author} {\bibfnamefont {P.}~\bibnamefont {Ao}}, \ and\
  \bibinfo {author} {\bibfnamefont {Q.}~\bibnamefont {Niu}},\ }\href {\doibase
  10.1103/PhysRevLett.76.3758} {\bibfield  {journal} {\bibinfo  {journal}
  {Phys. Rev. Lett.}\ }\textbf {\bibinfo {volume} {76}},\ \bibinfo {pages}
  {3758} (\bibinfo {year} {1996})}\BibitemShut {NoStop}%
\bibitem [{\citenamefont {Ao}\ and\ \citenamefont {Thouless}(1993)}]{ao1993}%
  \BibitemOpen
  \bibfield  {author} {\bibinfo {author} {\bibfnamefont {P.}~\bibnamefont
  {Ao}}\ and\ \bibinfo {author} {\bibfnamefont {D.~J.}\ \bibnamefont
  {Thouless}},\ }\href {\doibase 10.1103/PhysRevLett.70.2158} {\bibfield
  {journal} {\bibinfo  {journal} {Phys. Rev. Lett.}\ }\textbf {\bibinfo
  {volume} {70}},\ \bibinfo {pages} {2158} (\bibinfo {year}
  {1993})}\BibitemShut {NoStop}%
\bibitem [{\citenamefont {Sonin}(1997)}]{sonin1997}%
  \BibitemOpen
  \bibfield  {author} {\bibinfo {author} {\bibfnamefont {E.~B.}\ \bibnamefont
  {Sonin}},\ }\href {\doibase 10.1103/PhysRevB.55.485} {\bibfield  {journal}
  {\bibinfo  {journal} {Phys. Rev. B}\ }\textbf {\bibinfo {volume} {55}},\
  \bibinfo {pages} {485} (\bibinfo {year} {1997})}\BibitemShut {NoStop}%
\bibitem [{\citenamefont {Simula}(2018)}]{simula2018}%
  \BibitemOpen
  \bibfield  {author} {\bibinfo {author} {\bibfnamefont {T.}~\bibnamefont
  {Simula}},\ }\href {\doibase 10.1103/PhysRevA.97.023609} {\bibfield
  {journal} {\bibinfo  {journal} {Phys. Rev. A}\ }\textbf {\bibinfo {volume}
  {97}},\ \bibinfo {pages} {023609} (\bibinfo {year} {2018})}\BibitemShut
  {NoStop}%
\bibitem [{\citenamefont {Kittel}(2004)}]{kittel2004}%
  \BibitemOpen
  \bibfield  {author} {\bibinfo {author} {\bibfnamefont {C.}~\bibnamefont
  {Kittel}},\ }\href@noop {} {\emph {\bibinfo {title} {Introduction to solid
  state physics}}},\ \bibinfo {edition} {8th}\ ed.\ (\bibinfo  {publisher}
  {Wiley},\ \bibinfo {year} {2004})\BibitemShut {NoStop}%
\bibitem [{\citenamefont {Karl}\ and\ \citenamefont
  {Gasenzer}(2017)}]{karl2017}%
  \BibitemOpen
  \bibfield  {author} {\bibinfo {author} {\bibfnamefont {M.}~\bibnamefont
  {Karl}}\ and\ \bibinfo {author} {\bibfnamefont {T.}~\bibnamefont
  {Gasenzer}},\ }\href {https://doi.org/10.1088/1367-2630/aa7eeb} {\bibfield
  {journal} {\bibinfo  {journal} {New J. Phys.}\ }\textbf {\bibinfo {volume}
  {19}},\ \bibinfo {pages} {093014} (\bibinfo {year} {2017})}\BibitemShut
  {NoStop}%
\bibitem [{\citenamefont {Comaron}\ \emph {et~al.}(2019)\citenamefont
  {Comaron}, \citenamefont {Larcher}, \citenamefont {Dalfovo},\ and\
  \citenamefont {Proukakis}}]{comaron2019}%
  \BibitemOpen
  \bibfield  {author} {\bibinfo {author} {\bibfnamefont {P.}~\bibnamefont
  {Comaron}}, \bibinfo {author} {\bibfnamefont {F.}~\bibnamefont {Larcher}},
  \bibinfo {author} {\bibfnamefont {F.}~\bibnamefont {Dalfovo}}, \ and\
  \bibinfo {author} {\bibfnamefont {N.~P.}\ \bibnamefont {Proukakis}},\ }\href
  {\doibase 10.1103/PhysRevA.100.033618} {\bibfield  {journal} {\bibinfo
  {journal} {Phys. Rev. A}\ }\textbf {\bibinfo {volume} {100}},\ \bibinfo
  {pages} {033618} (\bibinfo {year} {2019})}\BibitemShut {NoStop}%
\bibitem [{\citenamefont {Groszek}\ \emph {et~al.}(2020)\citenamefont
  {Groszek}, \citenamefont {Comaron}, \citenamefont {Proukakis},\ and\
  \citenamefont {Billam}}]{groszek2020}%
  \BibitemOpen
  \bibfield  {author} {\bibinfo {author} {\bibfnamefont {A.~J.}\ \bibnamefont
  {Groszek}}, \bibinfo {author} {\bibfnamefont {P.}~\bibnamefont {Comaron}},
  \bibinfo {author} {\bibfnamefont {N.~P.}\ \bibnamefont {Proukakis}}, \ and\
  \bibinfo {author} {\bibfnamefont {T.~P.}\ \bibnamefont {Billam}},\
  }\href@noop {} {\bibfield  {journal} {\bibinfo  {journal} {arXiv:2006.13284}\
  } (\bibinfo {year} {2020})}\BibitemShut {NoStop}%
\bibitem [{\citenamefont {Williamson}\ and\ \citenamefont
  {Blakie}(2017)}]{williamson2017}%
  \BibitemOpen
  \bibfield  {author} {\bibinfo {author} {\bibfnamefont {L.~A.}\ \bibnamefont
  {Williamson}}\ and\ \bibinfo {author} {\bibfnamefont {P.~B.}\ \bibnamefont
  {Blakie}},\ }\href {\doibase 10.1103/PhysRevLett.119.255301} {\bibfield
  {journal} {\bibinfo  {journal} {Phys. Rev. Lett.}\ }\textbf {\bibinfo
  {volume} {119}},\ \bibinfo {pages} {255301} (\bibinfo {year}
  {2017})}\BibitemShut {NoStop}%
\bibitem [{\citenamefont {Geldhauser}\ and\ \citenamefont
  {Romito}(2019)}]{geldhauser2019}%
  \BibitemOpen
  \bibfield  {author} {\bibinfo {author} {\bibfnamefont {C.}~\bibnamefont
  {Geldhauser}}\ and\ \bibinfo {author} {\bibfnamefont {M.}~\bibnamefont
  {Romito}},\ }\href {\doibase 10.3934/math.2019.3.534} {\bibfield  {journal}
  {\bibinfo  {journal} {AIMS Mathematics}\ }\textbf {\bibinfo {volume} {4}},\
  \bibinfo {pages} {534} (\bibinfo {year} {2019})}\BibitemShut {NoStop}%
\bibitem [{\citenamefont {Higbie}\ \emph {et~al.}(2005)\citenamefont {Higbie},
  \citenamefont {Sadler}, \citenamefont {Inouye}, \citenamefont {Chikkatur},
  \citenamefont {Leslie}, \citenamefont {Moore}, \citenamefont {Savalli},\ and\
  \citenamefont {Stamper-Kurn}}]{higbie2005}%
  \BibitemOpen
  \bibfield  {author} {\bibinfo {author} {\bibfnamefont {J.~M.}\ \bibnamefont
  {Higbie}}, \bibinfo {author} {\bibfnamefont {L.~E.}\ \bibnamefont {Sadler}},
  \bibinfo {author} {\bibfnamefont {S.}~\bibnamefont {Inouye}}, \bibinfo
  {author} {\bibfnamefont {A.~P.}\ \bibnamefont {Chikkatur}}, \bibinfo {author}
  {\bibfnamefont {S.~R.}\ \bibnamefont {Leslie}}, \bibinfo {author}
  {\bibfnamefont {K.~L.}\ \bibnamefont {Moore}}, \bibinfo {author}
  {\bibfnamefont {V.}~\bibnamefont {Savalli}}, \ and\ \bibinfo {author}
  {\bibfnamefont {D.~M.}\ \bibnamefont {Stamper-Kurn}},\ }\href {\doibase
  10.1103/PhysRevLett.95.050401} {\bibfield  {journal} {\bibinfo  {journal}
  {Phys. Rev. Lett.}\ }\textbf {\bibinfo {volume} {95}},\ \bibinfo {pages}
  {050401} (\bibinfo {year} {2005})}\BibitemShut {NoStop}%
\bibitem [{\citenamefont {Muessel}\ \emph {et~al.}(2013)\citenamefont
  {Muessel}, \citenamefont {Strobel}, \citenamefont {Joos}, \citenamefont
  {Nicklas}, \citenamefont {Stroescu}, \citenamefont {Tomkovi{\v{c}}},
  \citenamefont {Hume},\ and\ \citenamefont {Oberthaler}}]{muessel2013}%
  \BibitemOpen
  \bibfield  {author} {\bibinfo {author} {\bibfnamefont {W.}~\bibnamefont
  {Muessel}}, \bibinfo {author} {\bibfnamefont {H.}~\bibnamefont {Strobel}},
  \bibinfo {author} {\bibfnamefont {M.}~\bibnamefont {Joos}}, \bibinfo {author}
  {\bibfnamefont {E.}~\bibnamefont {Nicklas}}, \bibinfo {author} {\bibfnamefont
  {I.}~\bibnamefont {Stroescu}}, \bibinfo {author} {\bibfnamefont
  {J.}~\bibnamefont {Tomkovi{\v{c}}}}, \bibinfo {author} {\bibfnamefont
  {D.~B.}\ \bibnamefont {Hume}}, \ and\ \bibinfo {author} {\bibfnamefont
  {M.~K.}\ \bibnamefont {Oberthaler}},\ }\href
  {https://doi.org/10.1007/s00340-013-5553-8} {\bibfield  {journal} {\bibinfo
  {journal} {Appl. Phys. B}\ }\textbf {\bibinfo {volume} {113}},\ \bibinfo
  {pages} {69} (\bibinfo {year} {2013})}\BibitemShut {NoStop}%
\bibitem [{\citenamefont {Pr{\"u}fer}\ \emph {et~al.}(2018)\citenamefont
  {Pr{\"u}fer}, \citenamefont {Kunkel}, \citenamefont {Strobel}, \citenamefont
  {Lannig}, \citenamefont {Linnemann}, \citenamefont {Schmied}, \citenamefont
  {Berges}, \citenamefont {Gasenzer},\ and\ \citenamefont
  {Oberthaler}}]{prufer2018}%
  \BibitemOpen
  \bibfield  {author} {\bibinfo {author} {\bibfnamefont {M.}~\bibnamefont
  {Pr{\"u}fer}}, \bibinfo {author} {\bibfnamefont {P.}~\bibnamefont {Kunkel}},
  \bibinfo {author} {\bibfnamefont {H.}~\bibnamefont {Strobel}}, \bibinfo
  {author} {\bibfnamefont {S.}~\bibnamefont {Lannig}}, \bibinfo {author}
  {\bibfnamefont {D.}~\bibnamefont {Linnemann}}, \bibinfo {author}
  {\bibfnamefont {C.-M.}\ \bibnamefont {Schmied}}, \bibinfo {author}
  {\bibfnamefont {J.}~\bibnamefont {Berges}}, \bibinfo {author} {\bibfnamefont
  {T.}~\bibnamefont {Gasenzer}}, \ and\ \bibinfo {author} {\bibfnamefont
  {M.~K.}\ \bibnamefont {Oberthaler}},\ }\href
  {https://doi.org/10.1038/s41586-018-0659-0} {\bibfield  {journal} {\bibinfo
  {journal} {Nature}\ }\textbf {\bibinfo {volume} {563}},\ \bibinfo {pages}
  {217} (\bibinfo {year} {2018})}\BibitemShut {NoStop}%
\bibitem [{\citenamefont {Wiegmann}\ and\ \citenamefont
  {Abanov}(2014)}]{wiegmann2014}%
  \BibitemOpen
  \bibfield  {author} {\bibinfo {author} {\bibfnamefont {P.}~\bibnamefont
  {Wiegmann}}\ and\ \bibinfo {author} {\bibfnamefont {A.~G.}\ \bibnamefont
  {Abanov}},\ }\href {\doibase 10.1103/PhysRevLett.113.034501} {\bibfield
  {journal} {\bibinfo  {journal} {Phys. Rev. Lett.}\ }\textbf {\bibinfo
  {volume} {113}},\ \bibinfo {pages} {034501} (\bibinfo {year}
  {2014})}\BibitemShut {NoStop}%
\bibitem [{\citenamefont {Yu}\ and\ \citenamefont
  {Bradley}(2017)}]{xiaoquan2017}%
  \BibitemOpen
  \bibfield  {author} {\bibinfo {author} {\bibfnamefont {X.}~\bibnamefont
  {Yu}}\ and\ \bibinfo {author} {\bibfnamefont {A.~S.}\ \bibnamefont
  {Bradley}},\ }\href {\doibase 10.1103/PhysRevLett.119.185301} {\bibfield
  {journal} {\bibinfo  {journal} {Phys. Rev. Lett.}\ }\textbf {\bibinfo
  {volume} {119}},\ \bibinfo {pages} {185301} (\bibinfo {year}
  {2017})}\BibitemShut {NoStop}%
\end{thebibliography}

%

\end{document}